\documentclass{aastex631}

\usepackage{caption}
\usepackage{subcaption}
\usepackage{bm}
\usepackage{amsmath}
\usepackage{multirow}
\usepackage{makecell}
\usepackage{array}
\DeclareMathOperator\erfc{erfc}

\begin{document}

\title{Disentangling $\gamma-\beta$: the 4$^{\rm th}$-order velocity moments based on spherical Jeans analysis}

\author{Dafa Wardana}
\email{dafaward@astr.tohoku.ac.jp}
\affiliation{Astronomical Institute, Tohoku University, 6-3 Aoba, Sendai, Japan}

\author{Masashi Chiba}
\affiliation{Astronomical Institute, Tohoku University, 6-3 Aoba, Sendai, Japan}

\author{Kohei Hayashi}
\affiliation{National Institute of Technology, Sendai College, 48 Nodayama, Natori, Japan}
\affiliation{Astronomical Institute, Tohoku University, 6-3 Aoba, Sendai, Japan}
\affiliation{ICRR, The University of Tokyo, 5-1-5 Kashiwanoha, Kashiwa, Japan}



\begin{abstract}
Distinguishing a core and a cusp within dark matter halos is complexified by the existence of mass-anisotropy degeneracy, where various combinations of velocity anisotropy ($\beta$) and inner density slope ($\gamma$) yield similar observational signatures. We construct a dynamical model that incorporates the 4th-order velocity moments to alleviate this challenge. The inclusion of the 4th-order velocity moments enables stars’ line-of-sight velocity distribution (LOSVD) to be flexible. This flexible LOSVD can cover from a thin-tailed to a heavy-tailed distribution that is inaccessible if only the 2nd-order moments are considered. We test the model on four mock galaxies having isotropic orbits, $\beta = 0$: two resembling dwarf spheroidal galaxies (dSphs) and two resembling ultra-faint dwarfs (UFDs) in terms of velocity dispersion. Each category includes one galaxy with a cuspy NFW profile and one with a cored density profile. Results show that a ratio of the global velocity dispersion to velocity error, $\sigma_{\rm los,global} / \delta v_{\rm los} \gtrsim 4$, is crucial to avoid systematic biases arising from the strong sensitivity of 4th-order moments to the LOSVD tails. In cases where this velocity ratio condition is met, our model reliably recovers $\gamma$ in dSph mock galaxies, with the true value recovered within $\sim 1\sigma$, and strongly excludes a cuspy NFW profile for the cored dSph mock galaxy. However, recovering the density profiles of UFDs remains challenging due to their intrinsically low velocity dispersions.

\end{abstract}



\section{Introduction} \label{sec:intro}

The $\Lambda$CDM theory has been facing some challenges, specifically in the subgalactic scales, with the core-cusp problem \citep{Flores1994,Moore1994,Burkert1995} being one of the oldest (see \citet{BullockBoylan-Kolchin2017} and \citet{Salucci2019} for reviews). This problem signifies a discrepancy in the central density slope of dark matter halos as inferred from simulations involving only dark matter particles compared to those inferred from the analysis of observational data. Dark matter-only simulations predict the universality of the dark matter density profile that rises steeply in the inner part following $\rho(r) \propto r^{-\gamma}$, with $\gamma \approx 1.0$ regardless the initial condition, well known as the NFW profile \citep{Navarro1996}. This universal property is in tension with the results from kinematics analyses using observational data from various types of dwarf galaxies that do not converge and, instead, favor a diverse inner density slope where $0 \lesssim \gamma \lesssim 1.5$ (e.g., \citet{Oh2015}).

While the existence of the core profile remains a subject of debate and is potentially constrained by modeling limitations \citep{deBlok2010}, the focus of probing a dark matter density profile has been redirected from gas-rich dwarf spirals \citep{Burkert1995, Salucci2000} to the nearby region; satellites of the Milky Way. The Milky Way dwarf spheroidal (dSph) and ultra-faint dwarf (UFD) satellites are among the most dark matter-dominated galaxies known at the moment \citep{Simon2019,BattagliaNipoti2022}. These types of galaxies are believed to be the building blocks of the MW-sized galaxies and retain the most pristine relics of ancient structures from the early universe. Their unbeatable low fraction of baryon content, even inside the half-light radius, makes them the most attractive sites to test dark matter theories because the baryonic effects are less significant compared to galaxies with a higher fraction of baryonic component. Their proximity also offers a key advantage that enables their member stars to be resolved, allowing for the line-of-sight velocity, projected position, and metallicity of the member stars to be obtained individually.

Modeling the internal dynamics of dSphs and UFDs has been considered one of the primary pathways in addressing the core-cusp problem (see \citet{BattagliaHelmiBreddels2013} for a review). Despite the advantages mentioned earlier, distinguishing between a core and a cusp remains challenging. Obtained results are often unable to robustly discriminate the two profiles from each other based on the available data set \citep{Breddels2013,Sameth2015}.
One central problem is the mass-anisotropy degeneracy, where a wide range of models with varying velocity anisotropies and density profiles can reproduce the observed kinematics. Improvements from various directions are made to break this degeneracy. 

Where available, the use of more than one chemodynamically distinct stellar population can partially alleviate the degeneracy \citep{WalkerPenarrubia2011,AmoriscoEvans2012StellarPopulations,Klaudia2022}. Motivated by the elongated observed stellar distribution and non-sphericity of simulated dark halo, there are also attempts to explore axisymmetric models, investigating different kinematical signatures along the major and minor axes \citep{HayashiChiba2012,HayashiChiba2015,HayashiChibaIshiyama2020}. Other groups aim to mitigate this degeneracy by developing Schwarzschild orbital-based superposition methods, assuming either a spherically symmetric \citep{Breddels2013,Klaudia2019} or axisymmetric \citep{Jardel2012,Jardel2013,Hagen2019} underlying gravitational potential.
In addition, methods based on distribution functions (DFs) offer a robust framework for dynamical modeling. These DF-based methods, as developed by \citet{AmoriscoEvans2011}, \citet{Pascale2018}, and \citet{Strigari2017}, ensure self-consistency by directly modeling the phase-space distribution of stars. By satisfying the collisionless Boltzmann equation, these methods naturally reproduce the observed line-of-sight velocity distributions (LOSVDs).
Another approach involves utilizing higher-order moments through the virial shape parameter \citep{RichardsonFairbairn2014,Genina2020ToBOrNotToB} or Jeans modeling \citep{Lokas2002,Lokas2009,RichardsonFairbairn2013}.

The limited exploration through the higher-order moments primarily stems from their heavy reliance on stars in the velocity tails and large sample size requirements \citep{Merrifield1990}.
The analysis involving higher-order moments is further complicated by the substantial impact of interlopers, which can significantly alter the velocity distribution in the tails where the star count is low.
Nevertheless, this situation is going to be improved in the near future, thanks to the Subaru Prime Focus Spectrograph project that will increase the kinematic sample in each of the targeted MW's dwarf galaxies to approximately 5,000 stars with expected velocity error measurement of $\sim$2 km~s$^{-1}$ \citep{Takada2014,Tamura2016,hayashi2023b}.
Therefore, exploring the ability and behavior of the higher-order moments to mitigate mass-anisotropy degeneracy becomes a subject of investigation.
In this work, we construct a dynamical model based on the spherically symmetric 2nd-order combined with 4th-order Jeans equations.

This paper is organized as follows. We initiate by detailing the dynamical model, which relies on the spherically symmetric 2nd-order and 4th-order Jeans equations, in Section 2. Section 3 presents the outcomes derived from applying the model to mock data. In section 4, we discuss the recovery of $\gamma$ and the recovery of $\beta$. We summarize our main conclusions in section 5.
\section{Models} \label{sec:models}
In this section, we provide a detailed overview of the components and methods used in our analysis. First, we describe the adopted distributions for dark matter and stars (Section \ref{subsec:dmdensity}). Next, we outline the 2nd-order (Section \ref{subsec:2ndjeans}) and 4th-order (Section \ref{subsec:4thjeans}) Jeans modeling frameworks employed in this study. We then explain the parameter marginalization process in the fitting procedure (Section \ref{subsec:fitting}), followed by a description of the mock data used for testing and validation (Section \ref{subsec:mockdata}).

Due to the dark matter domination even in the inner part of the galaxy, we assume that stars only act as tracers inside a gravitational potential provided by dark matter. We also assume that the effects of binaries are removed from the l.o.s. velocity dispersion and kurtosis profile and consider the case of a perfect determination of stars’ membership.

\subsection{Dark matter density and light profiles}
\label{subsec:dmdensity}
For the dark matter density profile, we adopt the Generalized Hernquist model introduced by \citet{Hernquist1990} and \citet{Zao1996}
\begin{equation}
    \rho_{\rm dm}(r) = \rho_0 \left( \frac{r}{a_{\rm dm}} \right)^{-\gamma} \left[ 1 + \left( \frac{r}{a_{\rm dm}} \right)^{\beta_{\rm trans}} \right]^{- ({\alpha-\gamma})/{\beta_{\rm  trans}}}.
    \label{GHM}
\end{equation}
In Equation (\ref{GHM}), $\rho_0$ is the dark matter scale density, $a_{\rm dm}$ is the dark matter scale length, $\gamma$ is the inner density slope, and $\beta_{\rm trans}$ is the transition sharpness between the inner and the outer slope density profile. A smaller value of $\beta_{\rm trans}$ means a smoother, more gradual change from the inner slope to the outer slope. The last parameter, $\alpha$, is the outer slope density profile that dominates the slope of the density profile in the galactocentric distance larger than $a_{\rm dm}$. This model provides a wide range of density profiles, including the cored profile ($\gamma \approx 0$), the cuspy profile ($\gamma \approx 1$), and any possible profiles between the two.

For the light profile, we adopt the Plummer model \citep{Plummer1911} that is given by
\begin{equation}
    \nu (r) = \frac{3L}{4\pi a_*^3} \left(1+ \frac{r^2}{a_*^2} \right)^{-5/2},
\end{equation}
where $\nu(r)$ is the 3-dimension light profile, $L$ is the total luminosity of the stellar system, and $a_*$ is a Plummer scale length, which in this case, also corresponds to the projected half-light radius. The observable projected light profile $I(R)$, where $R$ is the projected radius on the sky plane, can be obtained by integrating along the line-of-sight, which results
\begin{equation}
    I(R) = \frac{L}{\pi a_*^2} \left(1+ \frac{R^2}{a_*^2} \right)^{-2}.
\end{equation}
For simplicity, we introduce a dimensionless parameter $b\equiv a_{\rm dm}/a_*$.

\subsection{Spherically symmetric 2$^{\rm nd}$-order Jeans equations}
\label{subsec:2ndjeans}
Assuming that a gravitational system is static and spherical, the motion of its stars moving under the influence of underlying gravitational potential $\Phi(r)$ can be explained by the distribution function $f(\textbf{r},\textbf{v})$. However, the distribution function is inaccessible in the real application. Jeans equations link the inaccessible distribution function to the gravitational potential via velocity moments. We define the velocity moments in spherical coordinates ($r, \theta, \phi$) as:
\begin{equation}
    \nu \overline{v_r^i v_\theta^j v_\phi^k} = \int \mathrm{d}^3 v v_r^i v_\theta^j v_\phi^k f.
    \label{eqn:moments}
\end{equation}
For the second order, velocity moments represent the velocity dispersions $\overline{v_r^2} = \sigma_r^2$, $\overline{v_\theta^2} = \sigma_\theta^2$, and $\overline{v_\phi^2} = \sigma_\phi^2$, where the streaming motions are assumed to be zero. The 2nd-order Jeans equations can be written as \citep{BinneyTremaine2008}
\begin{equation}
    \frac{\mathrm{d}}{\mathrm{d} r} \nu \sigma_r^2 + 2 \frac{\beta}{r} \nu \sigma_r^2 + \nu \frac{\mathrm{d} \Phi}{\mathrm{d} r}= 0,
    \label{eqn:Jeans2nd1}
\end{equation}
where the velocity anisotropy $\beta$ is defined as $\beta \equiv 1 - (\sigma_\theta^2 + \sigma_\phi^2)/2\sigma_r^2$. In the case of constant $\beta$, i.e., assuming the distribution function is in the form of $f(E,L)=f_0(E)L^{-2\beta}$, where $E$ and $L$ denote the total energy and angular momentum, respectively, the solution of Equation (\ref{eqn:Jeans2nd1}) that satisfies the boundary condition $\lim_{r \to \infty} \sigma_r^2 = 0$ is
\begin{equation}
    \sigma_r^2 (r) = \frac{1}{r^{2\beta} \nu(r)} \int_r^\infty \mathrm{d} r' r'^{2\beta} \nu(r') \frac{\mathrm{d} \Phi}{\mathrm{d} r'},
    \label{sigmarr}
\end{equation}
The observable line-of-sight velocity dispersion $\sigma_{\rm los}$ as a function of projected radius, $R$, can be obtained by integrating along the line-of-sight
\begin{equation}
    \sigma^2_{\rm los}(R) = \frac{2}{I(R)} \int_R^\infty \mathrm{d} r \left( 1- \beta \frac{R^2}{r^2} \right) \frac{\nu \sigma_r^2 r}{\sqrt{r^2 -R^2}}.
    \label{sigmalos0}
\end{equation}

\subsection{Spherically symmetric 4$^{\rm th}$-order Jeans equations}
\label{subsec:4thjeans}
\begin{figure}
\centering
\includegraphics[width=0.5\textwidth]{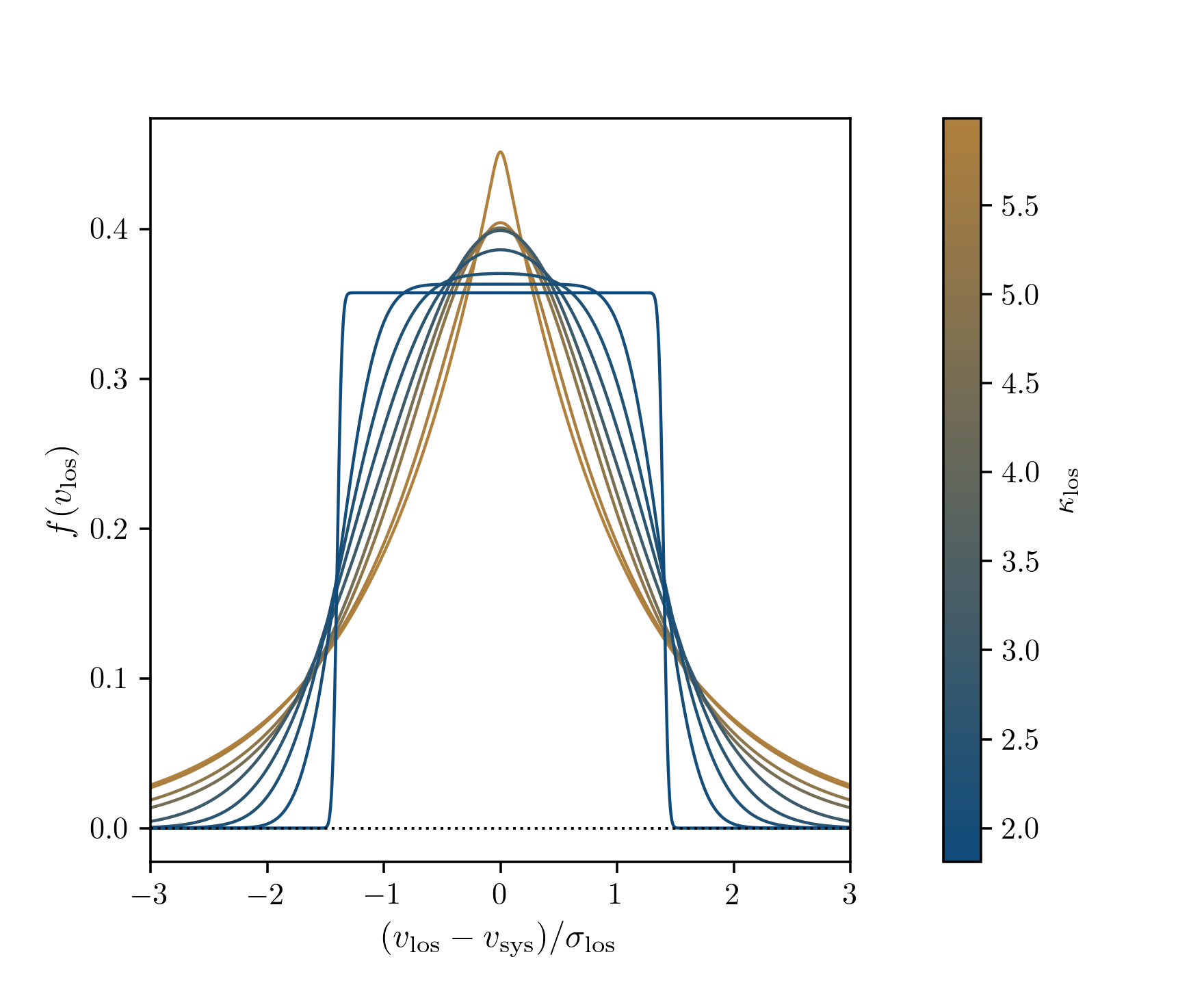}
\caption{
Probability density functions based on uniform and Laplacian kernels for scaled line-of-sight velocities across various choices of $\kappa_{\rm los}$.
}
\label{fig:kappaLOSVD}
\end{figure}

It is important to note that LOSVDs are generally non-Gaussian even in an isotropic system where the 3-dimensional velocity distributions are Gaussian everywhere \citep{BinneyMerrifield1998}. Hence, it is insufficient to describe LOSVDs only by the velocity dispersion alone as this case means that a non-Gaussian LOSVD is forced to be fit using the closest Gaussian distribution (see \citet{Read2021} for a demonstration). One needs other quantities that describe the degree of deviation from Gaussianity. While the 3rd-order moments correspond to the asymmetric deviation (skewness), the 4th-order moments describe the symmetric deviation from Gaussian distribution. Again, assuming a constant velocity anisotropy, the equation can be written as
\begin{equation}
    \frac{\mathrm{d}}{\mathrm{d} r} \nu \overline{v_r^4} + \frac{2\beta}{r} \nu \overline{v_r^4} + 3\nu \sigma_r^2 \frac{\mathrm{d} \Phi}{\mathrm{d} r}=0,
\end{equation}
whose solution may be expressed as
\begin{equation}
    \overline{ v_{r}^4}(r)  = \frac{3}{r^{2\beta}\nu(r)} \int_r^\infty \mathrm{d} r' r'^{2\beta} \nu(r') \sigma_r^2(r') \frac{\mathrm{d} \Phi}{\mathrm{d} r'}.
\end{equation}
The projected 4th-order moments take the form
\begin{equation}
    \overline{ v_{\rm los}^4}(R) = \frac{2}{I(R)} \int_R^\infty \mathrm{d} r \left[1-2\beta \frac{R^2}{r^2} +\frac{1}{2}\beta (1+\beta)\frac{R^4}{r^4} \right] \frac{\nu \overline{v_r^4}r}{\sqrt{r^2-R^2}}.
    \label{eqn:battaglia2013}
\end{equation}
It is convenient to express the 4th-order moments in the form of the line-of-sight kurtosis
\begin{equation}
    \kappa_{\rm los}(R) = \frac{\overline{v^4_{\rm los}}(R)}{\sigma_{\rm los}^4(R)},
    \label{eqn:kappa}
\end{equation}
\citep{Merrifield1990,Lokas2002} whose values are $\kappa_{\rm los} = 3$ for a Gaussian distribution, $\kappa_{\rm los} < 3$ for a more thin-tailed and flat-topped distribution, and $\kappa_{\rm los} > 3$ for a spikier and more heavy-tailed distribution (see Figure \ref{fig:kappaLOSVD}).

In the unbinned 2nd-order only Jeans modeling, one usually links the theoretical l.o.s. velocity dispersion $\sigma_{\rm los}$ to the LOSVD by assuming that its distribution is Gaussian. Together with the velocity error measurement $\delta v_{\rm los}$, which has a Gaussian distribution as well, these two quantities characterize the LOSVD for each member star at position $R$. However, taking into account the 4th-order moments, we need to provide more freedom to the proposed LOSVD to change its shape. For this purpose, we adopt two kernels (uniform and Laplacian kernels, see Appendix A for details) explained in \citet{SandersEvans2020} to link the theoretical l.o.s. velocity dispersion and kurtosis to the LOSVD. Figure \ref{fig:kappaLOSVD} shows how the value of kurtosis changes the shape of a distribution in this model. Another notable advantage of adopting these kernels is that the velocity error measurements for each star can be convolved analytically into the distribution. Therefore, stars with precise measurements do not lose their strong signature of kinematical information because of averaging the error or their l.o.s. velocity is unnecessary.

\subsection{Fitting Procedure}
\label{subsec:fitting}
This work avoids a binned analysis because the results suffer from a strong dependency on the choice of binning and, hence, are prone to bias. The binned analysis is also avoided because extracting the observed $\kappa_{\rm los}$ by binning the sample can introduce a potentially large error of $\gtrsim 20\%$ unless each bin contains $\gtrsim 150$ stars \citep{Merrifield1990}. Instead of employing binning to estimate the most likely values of the free parameters, we adopt a different approach. Specifically, we calculate the probability of each star having the l.o.s. velocity $v_{{\rm los,}i}$, given a set of parameters $\theta(\beta, \alpha, \beta_{\rm trans}, \gamma, \rho_0, b)$ that describe the LOSVD at the star's projected position $R_i$. The LOSVD's width is characterized by the l.o.s velocity dispersion $\sigma_{{\rm los,}i}(R_i|\theta)$ calculated through Equation (\ref{sigmalos0}) and convolved with the observational velocity error $\delta v_{\mathrm{los,}i}$. Meanwhile, the l.o.s kurtosis $\kappa_{\rm los}(R_i|\theta)$ that is calculated through Equation (\ref{eqn:kappa}) controls the LOSVD's shape.

We assume that there is no net streaming motion. Therefore, LOSVDs are symmetrically centered at the systemic velocity of the stellar system $v_{\rm sys}$.
For $N$ total number of sample stars in a stellar system, the likelihood function is given by
\begin{equation}
    L = \prod_{i=1}^{N} \frac{1}{ (\delta v_{\mathrm{los,}i}^2 + \sigma_{{\rm los,}i}^2)^{1/2}} f_s(w_i),
\end{equation}
where
\begin{equation}
w_i^2 = \frac{(v_{{\rm los,}i} - v_{\rm sys})^2}{\delta v_{\mathrm{los,}i}^2 + \sigma_{{\rm los,}i}^2},
\end{equation}
and the choice of $f_s(w)$ is determined by the value of $\kappa_{\rm los}$. If a combination of parameters results in $\kappa_{\rm los}<3$, the uniform kernel is used, and if $\kappa_{\rm los}>3$, the Laplacian kernel is applied. Additionally, $\kappa_{\rm los}$ is linked to the uniform kernel via Equations (\ref{eq:fsuniform}) and (\ref{eq:kappalosuniform}), and to the Laplacian kernel via Equations (\ref{eq:fslaplacian}) and (\ref{eq:kappaloslaplacian}). These equations ensure that for any given $\kappa_{\rm los}$, there exists a unique solution for the distribution $f_s(w)$.
These choices of kernels allow the shape of LOSVDs to be flexible, covering more thin-tailed to heavy-tailed distributions than a Gaussian, with the limitation being the pure uniform ($\kappa_{\rm los} = 1.8$) and pure Laplacian ($\kappa_{\rm los} = 6.0$) distribution at each extreme case. Detailed properties of these kernels are given in Appendix \ref{sec:kernels}.
A flat or log-flat prior is applied to marginalize the free parameters in these given ranges:
\begin{enumerate}
    \item $ -1.0 \leqslant -\log_{10}(1-\beta) \leqslant 1.0 $,
    \item $ 2.0 \leqslant \alpha \leqslant 10.0 $,
    \item $ 0 \leqslant \beta_{ \rm trans} \leqslant 3.0 $,
    \item $ -2.0 \leqslant \gamma \leqslant 2.0 $,
    \item $ -5.0 \leqslant \log_{10}(\rho_0) \leqslant 2.0 $, and
    \item $ -2.0 \leqslant \log_{10}(b) \leqslant 2.0 $.
\end{enumerate}
Even though it is unphysical to have a system with $\gamma < 0$, which means a 'hole' in the center, we do not adopt a more restrictive prior range on $\gamma$ that excludes the negative region. The detail concerning this selection is explained in Section \ref{sec:priorgammadependence}. For simplicity, we assume that the projected stellar distribution is well-estimated from photometric measurements. Therefore, no parameter describing the stellar distribution is treated as a free parameter in our analysis. An alternative approach that might be more accommodating for observational uncertainties in the stellar distribution is demonstrated by \citet{Chang2021}, where a two-layered fitting procedure is employed: the first step involves estimating the stellar distribution, followed by a second step to constrain the gravitational potential.

Finally, we use the Markov Chain Monte Carlo (MCMC) method within the Metropolis-Hastings framework \citep{Metropolis1953, Hastings1970} to derive the posterior distribution of the free parameters. For this analysis, we employ a custom-made code specifically developed for this work. Convergence of the MCMC chains is assessed through visual inspection. Our MCMC setup involves 20 walkers, each generating 25,000 samples for each realization, with the first 5,000 samples discarded in the burn-in process to eliminate the initial point effects. After post-processing, this results in a total of 400,000 samples for each realization. Lastly, posterior distributions are generated by averaging across all realizations within the same scenario.

\subsection{Mock data}
\label{subsec:mockdata}
We generate mock galaxies using \textsc{StarSampler}\footnote{https://github.com/maoshenl/StarSampler} \citep{Liu2019PhDT} to apply and test our constructed dynamical model.
\textsc{StarSampler} is a Python-based tool developed to create random samples from any stationary distribution function (DF) specified by the user. This DF represents the probability density of stellar positions and velocities within six-dimensional phase space. The module is versatile, enabling the sampling of DFs that describe stellar systems, whether they are self-gravitating or situated within an external potential created by other sources, which is a dark matter halo in this case. Two sampling techniques are available, allowing users to select based on their desired level of efficiency and accuracy: the rejection sampling method, which delivers exact results but may be computationally inefficient, and the importance sampling method, which is faster but provides an approximate solution.

In this study, we set all mock galaxies to be isotropic, $\beta = 0$,  for the sake of simplicity. The light profile of the galaxies is modeled using a Plummer profile, which is a specific case of the Generalized Hernquist Model characterized by $(\alpha_\ast, \beta_{\mathrm{trans},\ast}, \gamma_\ast) = (5, 2, 0)$. Here, $\alpha_\ast$, $\beta_{\mathrm{trans},\ast}$, and $\gamma_\ast$ represent the outer slope, transition sharpness, and inner slope of the stellar density profile, respectively. For the dark matter potential, we adopt two distinct density profiles: the first is an NFW profile defined by $(\alpha, \beta_\mathrm{trans}, \gamma) = (3, 1, 1)$, and the second is a cored NFW profile with $(\alpha, \beta_\mathrm{trans}, \gamma) = (3, 1, 0)$.
There are two galaxies generated for each density profile to cover two different regimes: dSphs and the less bright, and less massive UFDs \citep{Simon2019,Battaglia2022}. This is done by differing the input value of dark matter scale density $\rho_0$ and dark matter scale length $a_{\rm dm}$ while keeping the intrinsic velocity dispersion profiles to be similar.
All details of the mock galaxies are summarized in Table \ref{Tab:mockdata}.


\begin{table}[htbp]
  \centering
  \caption{Input parameters of the mock data.}
  \begin{tabular}{ccccc}
    \hline
    Parameter & Cored dSph & Cuspy dSph & Cored UFD & Cuspy UFD \\
    \hline
    $\beta$ & 0.0 & 0.0 & 0.0 & 0.0 \\
    $\alpha$ & 3.0 & 3.0 & 3.0 & 3.0 \\
    $\beta_{ \rm trans}$ & 1.0 & 1.0 & 1.0 & 1.0 \\
    $\gamma$ & 0.0 & 1.0 & 0.0 & 1.0 \\
    $\rho_0$/[M$_\odot$ pc$^{-3}$] & $0.6$ & $0.064$  & $0.064$ & $0.002$\\
    $b$ & 4.5 & 10 & 10 & 50 \\
    
    \hline
  \end{tabular}
  \label{Tab:mockdata}
\end{table}

\begin{figure}[h]
  \centering
  \includegraphics[width=1.0\linewidth]{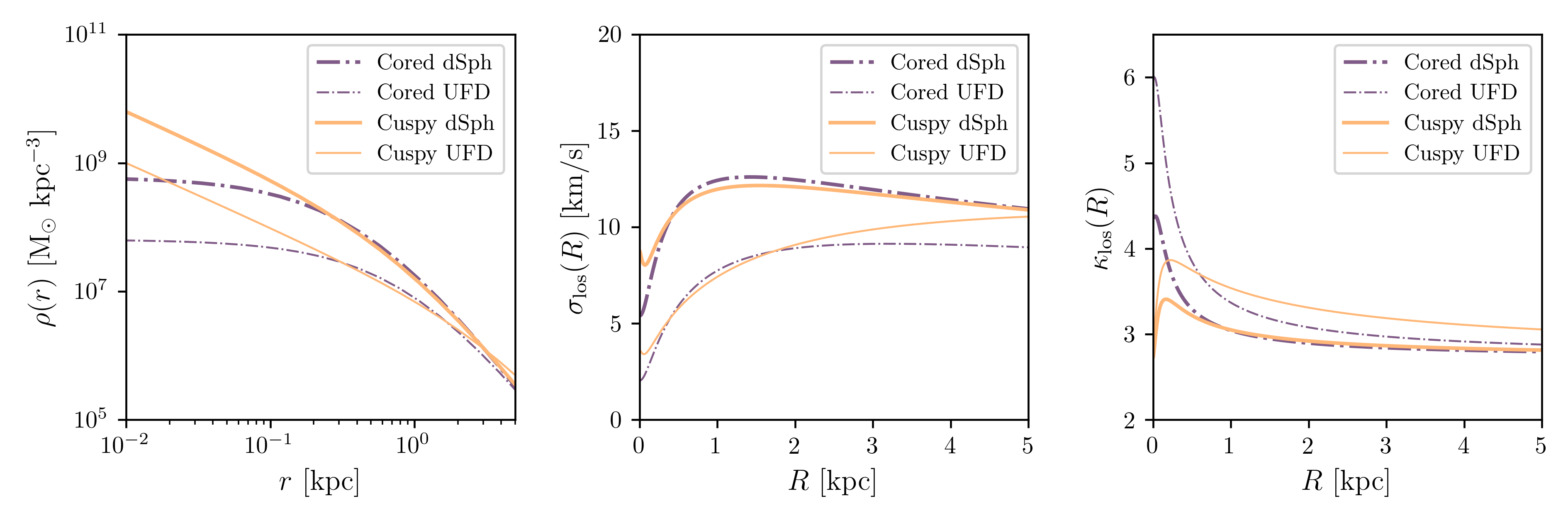}  
  \caption{Density, l.o.s. velocity dispersion, and l.o.s. kurtosis profiles for the mock galaxies are shown in the left, middle, and right panels, respectively. The purple dash-dotted lines represent galaxies with a cored density profile, while the orange solid lines indicate galaxies with a cuspy density profile. Mock galaxies resembling dSphs are shown with thicker lines, and those resembling UFDs are shown with thinner lines.}
  \label{fig:mockdatatrueprofiles}
\end{figure}

Figure \ref{fig:mockdatatrueprofiles} shows the intrinsic density profile (left panel), l.o.s. velocity dispersion (middle panel), and l.o.s. kurtosis (right panel) profile of the mock galaxies. Solid lines denote cuspy mock galaxy profiles, while dash-dotted lines denote the cored one. Thick lines mark mock galaxies resembling the classical MW's satellites, and the thinner lines are for mock galaxies similar to UFD's.

To ensure robust statistical analysis, we generate 10 realizations for each scenario set. These realizations are created for data sets containing 500 and 5,000 stars, respectively\footnote{The velocity dispersion calculated using 5,000 stars indicates that the dSph mock galaxies resemble Sextans and Leo II, while the UFD mock galaxies are comparable to Segue I and Reticulum II \citep{Battaglia2022}}. The choices of sample sizes are specifically designed to allow a comparison between the currently available data in dSphs and UFDs, and the expected data quality after the Subaru Prime Focus Spectrograph (PFS) survey \citep{Takada2014}. The independent results from each realization within the same scenario are then averaged to ensure consistency and mitigate the effects of statistical fluctuations. We assume two levels of measurement error in the line-of-sight velocity. Specifically, we set $\delta v_{\rm los} = 2$~km~s$^{-1}$ to represent the typical measurement error in current datasets and $\delta v_{\rm los} = 0.01$~km~s$^{-1}$ to suppress any biases associated with velocity uncertainties and identify any change present when we assign a larger one. For clarity and consistency, throughout this paper, we adopt the notation $R$ to represent the projected radius and $r$ to denote the three-dimensional galactocentric radius. This distinction ensures an unambiguous interpretation of results and parameters in the context of spherical symmetry.

\section{Results}
In this section, we present the results of the MCMC fitting conducted on the mock data. We examine the impact of varying the sample size, velocity measurement errors, and spatial distribution limits of the sampled stars. Our primary focus is on accurately recovering the inner density slope and velocity anisotropy, as well as exploring the potential degeneracy between these parameters. The complete posterior distributions for all free parameters are included in Appendix \ref{sec:6parsdistribution}. We begin by presenting the results derived from dSph mock galaxies (Section \ref{result:comparison} and \ref{result:observables}), followed by those obtained from UFD mock galaxies (Section \ref{result:UFD}).


\subsection{Comparison between the model with 4th-order moments \& 2nd-order moments only} \label{result:comparison}
\begin{figure}
  \centering
  \includegraphics[width=0.9\textwidth]{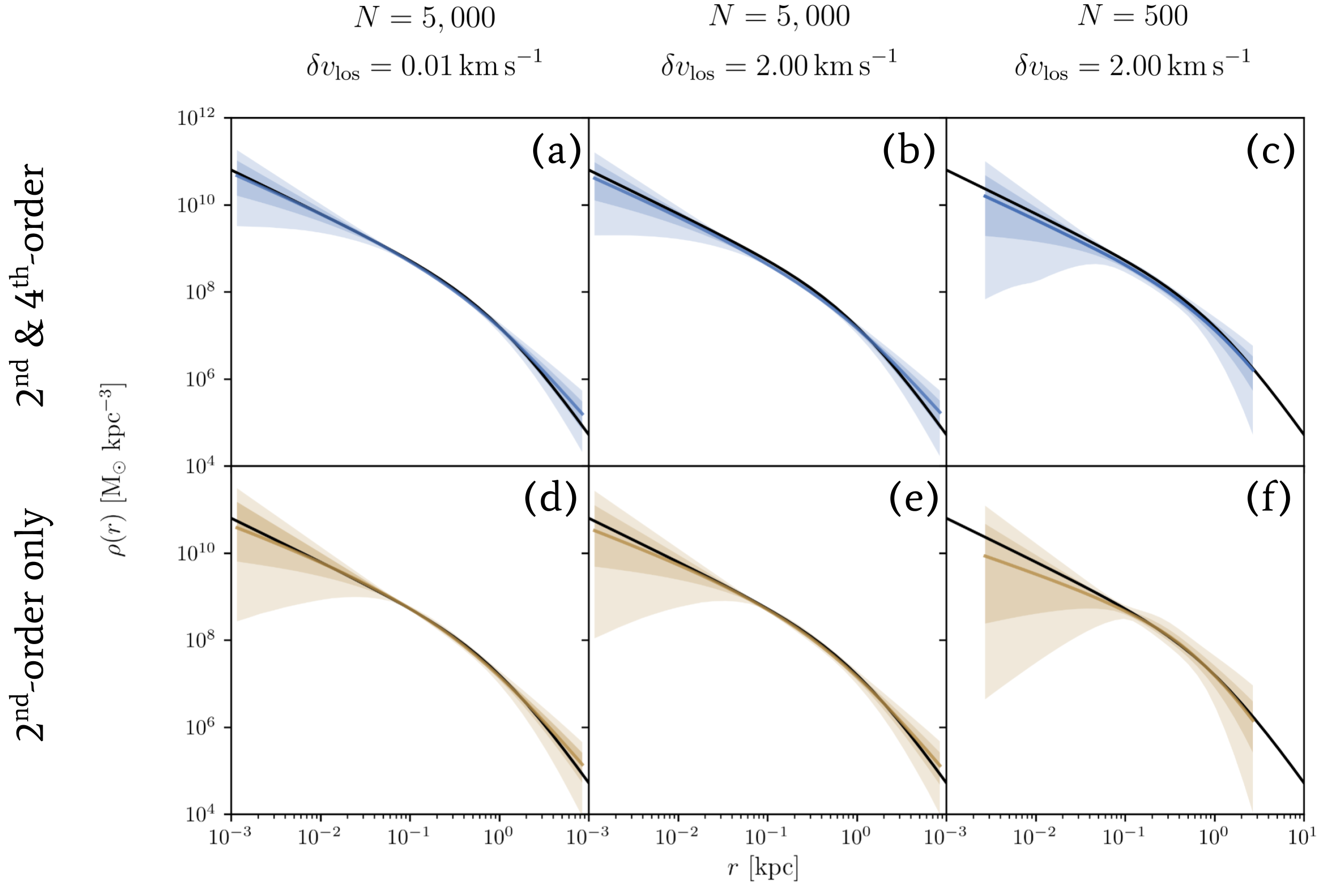}  
  \caption{Recovery of the dark matter density profile for the cuspy dSph mock galaxy. The left column [panel (a) and (d)] corresponds to fitting using 5,000 randomly selected stars with a uniform velocity error of 0.01~km~s$^{-1}$. The middle column [panel (b) and (e)] uses 5,000 stars with a uniform velocity error of 2~km~s$^{-1}$, while the right column [panel (c) and (f)] corresponds to 500 stars with a uniform velocity error of 2~km~s$^{-1}$. The upper panels with blue shades show results from modeling incorporating both the 2nd- and 4th-order velocity moments, while the lower panels with brown shades show results based on the 2nd-order velocity moments only. The solid black line represents the true dark matter density profile, different shaded regions show the 68\% and 95\% confidence intervals, and the solid blue or brown lines indicate the median recovered profiles. The half-light radius is located at $a_*=100$ pc in each panel.}
  \label{fig:densityprofrecovsphcusp}
\end{figure}

\begin{figure}
  \centering
  \includegraphics[width=0.9\textwidth]{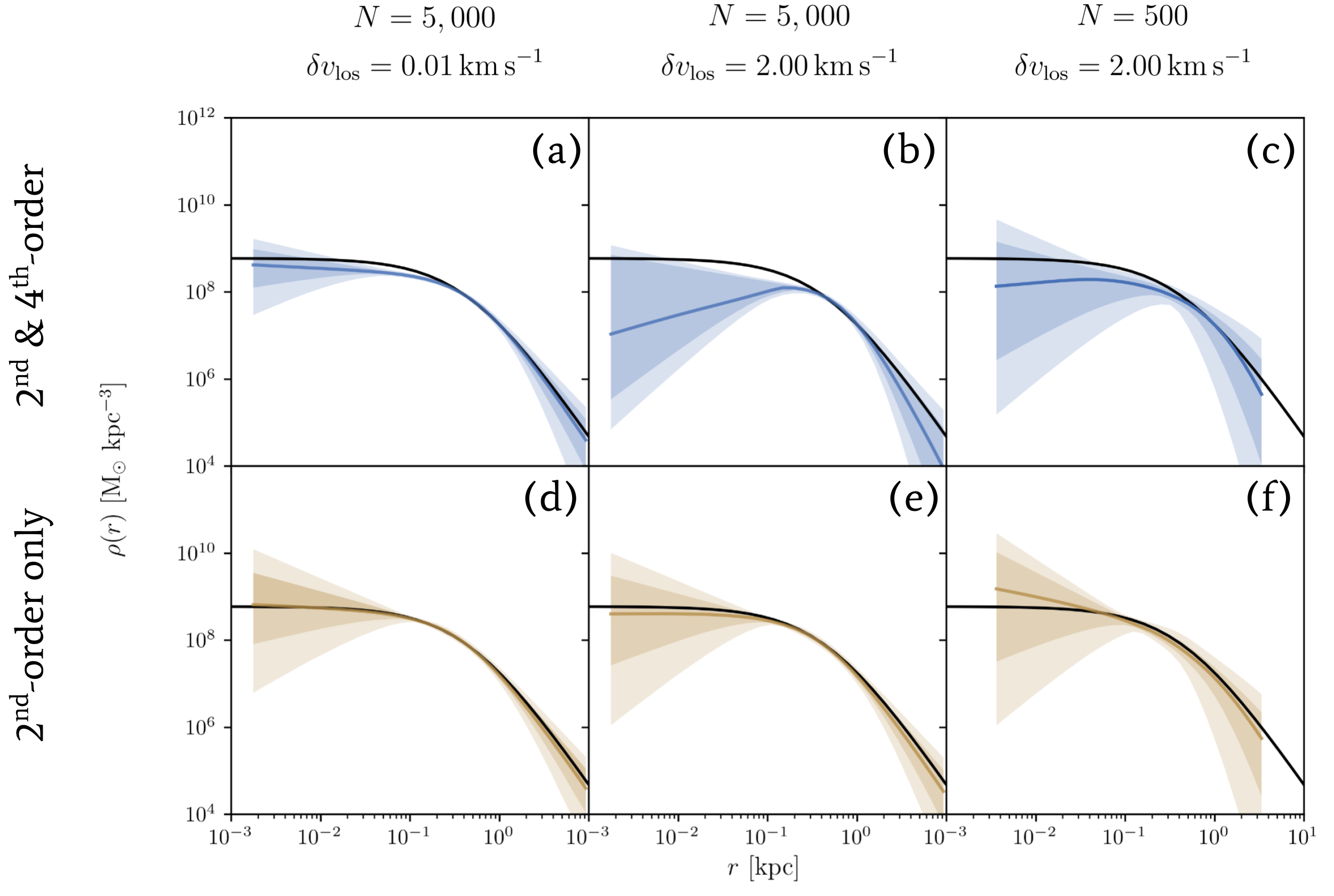}  
  \caption{The same as Figure \ref{fig:densityprofrecovsphcusp}, but for the cored dSph mock galaxy.}
  \label{fig:densityprofrecovsphcore}
\end{figure}

Figure \ref{fig:densityprofrecovsphcusp} presents the results of the density profile recovery in the range of $r$ where tracers are available for the cuspy dSph mock galaxy, while Figure \ref{fig:densityprofrecovsphcore} presents similar results for the cored dSph mock galaxy. These figures compare the performance of the dynamical model incorporating both 2nd- and 4th-order velocity moments (upper panels in blue) with the one that relies solely on 2nd-order moments (lower panels in brown). In both cases, the black lines denote the true underlying dark matter density profile, while the thick solid blue or brown lines represent the median recovered profiles. The shaded regions correspond to the 68\% and 95\% confidence intervals. Results in the left column correspond to fitting using 5,000 randomly selected stars with a uniform velocity error of 0.01~km~s$^{-1}$. In the middle column, we fit 5,000 stars with a uniform velocity error of 2~km~s$^{-1}$, while the right column corresponds to 500 stars with a uniform velocity error of 2~km~s$^{-1}$. The Plummer half-light radius is located at $a_* = 100$ pc in each panel.  Observational properties in the rightmost panels, in fact, resemble the typical currently available observational data set for most of the Milky Way's dSphs \citep{Mateo2008,Walker2009,Walker2009b,Walker2015,Fabrizio2016, Spencer2017,Spencer2018}. Meanwhile the middle panels are similar to what is expected for dSph targets after the Subaru PFS survey is performed \citep{Takada2014}. 

\begin{figure}
  \centering
  \includegraphics[width=1.0\textwidth]{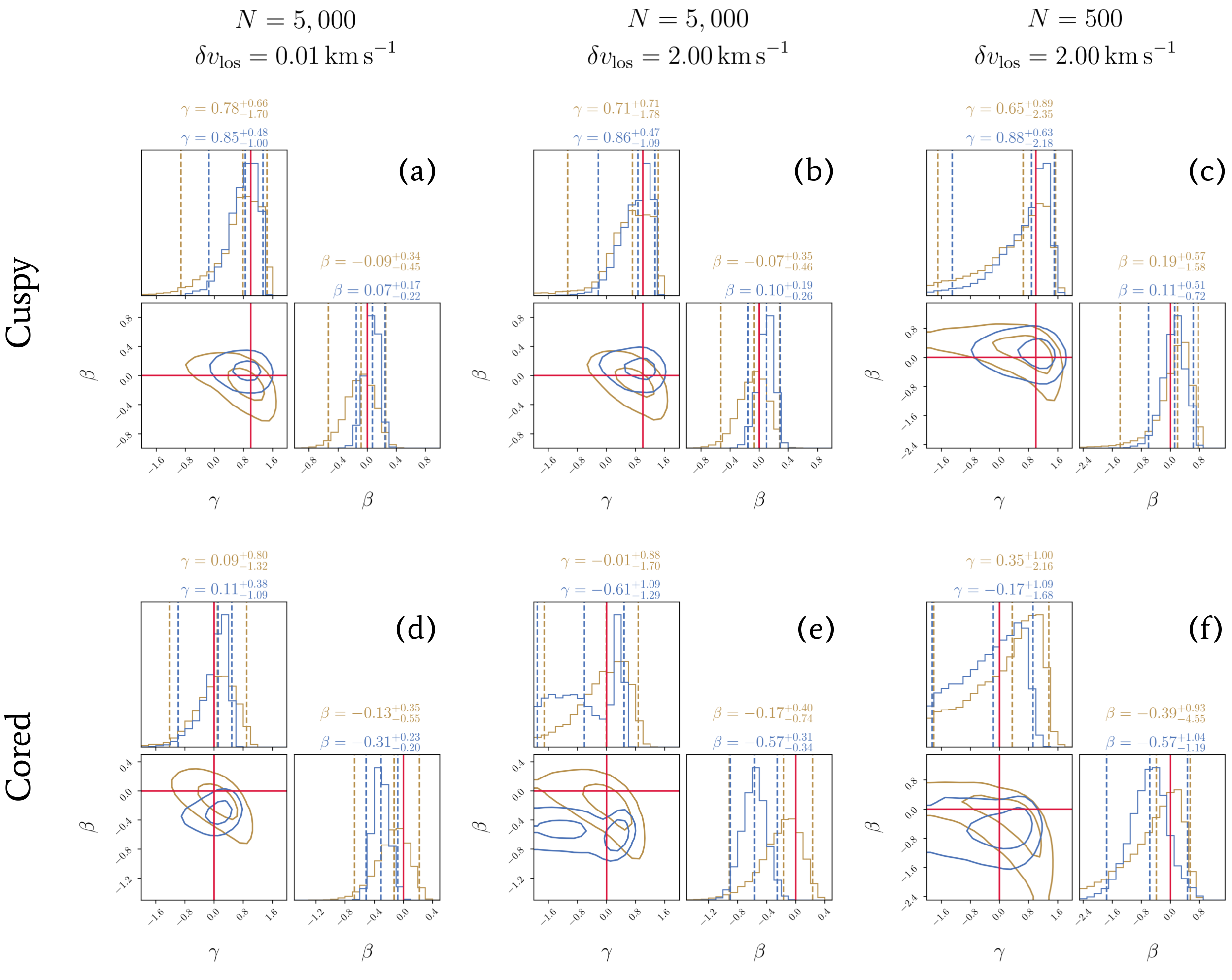}  
  \caption{ Posterior distributions for $\gamma$ and $\beta$ obtained from the fittings of a cuspy dSph (upper panels) and a cored dSph (lower panels). The brown represents the model relying solely on the 2nd-order moment, while the blue corresponds to the model incorporating both the 2nd-order and 4th-order moments. The left column corresponds to fitting using 5,000 randomly selected stars with a uniform velocity error of 0.01~km~s$^{-1}$. The middle column uses 5,000 stars with a uniform velocity error of 2~km~s$^{-1}$, while the right column corresponds to 500 stars with a uniform velocity error of 2~km~s$^{-1}$. Red solid lines denote the true values of each parameter, while the three vertical dashed lines indicate the median and 95\% confidence intervals (the 68\% confidence interval is presented in the Appendix to avoid clutter and improve visibility in this figure). Estimated median values and 95\% confidence intervals for each case are also indicated above the corresponding columns. Contour lines in the 2-dimensional posterior distributions represent the 68\% and 95\% confidence levels.}
  \label{fig:cornerplot_beta_gamma}
\end{figure}
In the cuspy dSph mock galaxy, the recovery of the density profile appears to be consistent across all panels. In each case, the true density profile is successfully recovered within the 68\% confidence level, with the associated uncertainties decreasing progressively from the rightmost to the leftmost panels. This trend is expected due to the improvement in data quality or quantity as we move leftward. However, using only the 2nd-order moments in the model reveals its limitations. Specifically, the model fails to rule out, within the 95\% confidence level, density profiles that decrease toward the center, which is unphysical. With 4th-order moments, such unrealistic density profiles are only present in panel (c), where the data quality is limited to 500 stars with a velocity error of 2~km~s$^{-1}$.

In contrast, the density profile recovery in the cored dSph case displays more variability than in the cuspy case. Panels (a) and (d) demonstrate that both models—using 2nd-order and 4th-order moments—can trace the cored density profile accurately. The model incorporating 4th-order moments performs particularly well, ruling out the cuspy profile entirely within the 95\% confidence level. However, the model relying solely on 2nd-order moments still exhibits broader uncertainty under the same conditions. Moving to panels (b) and (e), a dramatic shift emerges. The 4th-order moment model, which previously outperformed the 2nd-order-only model, fails to recover the inner part of the density profile. On the other hand, the 2nd-order-only model proves to be more resistant to the increase in velocity error, maintaining nearly the same performance as in panel (c) but with broader uncertainty. This comparison highlights the vulnerability of dynamical models incorporating 4th-order moments to higher velocity errors, whereas the 2nd-order-only model exhibits greater resilience in this regard.

Figure \ref{fig:cornerplot_beta_gamma} shows the comparison of the posterior distribution of model fitting based on only 2nd-order moments (brown) to the one using both 2nd-order \& 4th-order moments (blue). The sequence from left to right is still the same: 5000 stars with $\delta v_{\rm los} = 0.01$~km~s$^{-1}$, 5000 stars with $\delta v_{\rm los} = 2.00$~km~s$^{-1}$, and 500 stars with $\delta v_{\rm los} = 2.00$~km~s$^{-1}$, respectively. The upper panels correspond to the cuspy dSph mock galaxy, and the lower panels correspond to the cored dSph mock galaxy. Red solid lines indicate the true value, and three vertical dashed lines mark the median and 95\% confidence level. In the 2-dimensional distribution, contours show the 68\% and 95\% confidence levels. Note that the scale ranges for $\gamma$ are all the same, but for $\beta$ in the rightmost panels, the range is three times the other panels.

Compared to the model that uses only the 2nd-order moments, incorporating the 4th-order moments reduces the $\gamma - \beta$ degeneracy, as indicated by the less pronounced diagonally elongated features in the 2-dimensional contours. This reduction becomes more prominent as the number of stars increases. The most notable improvement for $\gamma$ is seen in panel (d) of Figure \ref{fig:cornerplot_beta_gamma}, where the incorrect cuspy NFW profile can be ruled out with a confidence level exceeding 3$\sigma$, despite an offset in the recovery of $\beta$. In the cuspy case, although the cored density profile can only be ruled out at the 1$\sigma$ level, the model incorporating the 4th-order moments consistently outperforms the 2nd-order-only model. Remarkably, it successfully recovers the cuspy profile even with as few as 500 stars and a velocity error of $2.00$~km~s$^{-1}$. We summarize the median values and 68\%  confidence intervals from dSph mock galaxy fittings in Table \ref{Tab:sphcuspresult} and \ref{Tab:sphcoredresult}. Columns labeled "2nd" present the results derived from the model that relies solely on the 2nd-order moments, while those labeled "4th" display the results obtained from the model incorporating both the 2nd- and 4th-order moments. In Table \ref{Tab:mediandiffsph}, we also present the standard deviation of the median values of $\gamma$ and $\beta$ across realizations, which reflects the variation among different sample sets.
\begin{table}[htbp]
  \centering
  \renewcommand{\arraystretch}{1.3} 
  \caption{Median and 68\% confidence level of the estimated parameters for the cuspy dSph mock galaxy}
  \begin{tabular}{cccccccc}
    \hline
    \multirow{3}{*}{Parameter} & \multirow{3}{*}{True value} & \multicolumn{2}{c}{$N = 5,000$} & \multicolumn{2}{c}{$N = 5,000$} & \multicolumn{2}{c}{$N = 500$} \\
                               &            & \multicolumn{2}{c}{$\delta v_{\rm los} = 0.01$ km s$^{-1}$} & \multicolumn{2}{c}{$\delta v_{\rm los} = 2.00$ km s$^{-1}$} & \multicolumn{2}{c}{$\delta v_{\rm los} = 2.00$ km s$^{-1}$} \\
                               &            & 2nd & 4th & 2nd & 4th & 2nd & 4th \\
    \hline
    $\beta$ & 0.00 & $-0.09_{-0.21}^{+0.18}$ & $0.07_{-0.11}^{+0.09}$ & $-0.07_{-0.21}^{+0.19}$ & $0.10_{-0.12}^{+0.10}$ & $0.19_{-0.50}^{+0.32}$ & $0.11_{-0.32}^{+0.28}$ \\
    $\alpha$ & 3.00 & $2.64_{-0.49}^{+0.94}$ & $2.68_{-0.50}^{+0.81}$ & $2.68_{-0.5}^{+0.96}$ & $2.58_{-0.42}^{+0.83}$ & $3.70_{-1.29}^{+2.72}$ & $3.36_{-1.05}^{+2.49}$ \\
    $\beta_{ \rm trans}$ & 1.00 & $0.84_{-0.37}^{+0.88}$ & $0.72_{-0.33}^{+0.84}$ & $0.79_{-0.35}^{+0.85}$ & $0.76_{-0.38}^{+1.15}$ & $1.39_{-0.78}^{+1.04}$ & $1.21_{-0.75}^{+1.17}$ \\
    $\gamma$ & 1.00 & $0.78_{-0.68}^{+0.45}$ & $0.85_{-0.44}^{+0.32}$ & $0.71_{-0.59}^{+0.50}$ & $0.86_{-0.51}^{+0.34}$ & $0.65_{-1.18}^{+0.61}$ & $0.88_{-0.87}^{+0.41}$ \\
    $\log(\rho_0$/[M$_\odot$ pc$^{-3}$]) & $-1.19$ & $-0.48_{-1.13}^{+1.08}$ & $-0.70_{-0.83}^{+1.15}$ & $-0.42_{-1.19}^{+1.07}$ & $-0.86_{-0.84}^{+1.34}$ & $-0.71_{-1.40}^{+1.33}$ & $-1.07_{-1.32}^{+1.67}$ \\
    $\log(b)$ & 1.00 & $0.76_{-0.60}^{+0.52}$ & $0.90_{-0.48}^{+0.42}$ & $0.79_{-0.57}^{+0.57}$ & $0.90_{-0.48}^{+0.48}$ & $1.01_{-0.65}^{+0.60}$ & $1.13_{-0.76}^{+0.55}$ \\
    
    \hline
  \end{tabular}
  \label{Tab:sphcuspresult}
\end{table}

\begin{table}[htbp]
  \centering
  \renewcommand{\arraystretch}{1.3} 
  \caption{Median and 68\% confidence level of the estimated parameters for the cored dSph mock galaxy}
  \begin{tabular}{cccccccc}
    \hline
    \multirow{3}{*}{Parameter} & \multirow{3}{*}{True value} & \multicolumn{2}{c}{$N = 5,000$} & \multicolumn{2}{c}{$N = 5,000$} & \multicolumn{2}{c}{$N = 500$} \\
                               &            & \multicolumn{2}{c}{$\delta v_{\rm los} = 0.01$ km s$^{-1}$} & \multicolumn{2}{c}{$\delta v_{\rm los} = 2.00$ km s$^{-1}$} & \multicolumn{2}{c}{$\delta v_{\rm los} = 2.00$ km s$^{-1}$} \\
                               &            & 2nd & 4th & 2nd & 4th & 2nd & 4th \\
    \hline
    $\beta$ & 0.00 & $-0.13_{-0.24}^{+0.20}$ & $-0.31_{-0.11}^{+0.12}$ & $-0.17_{-0.31}^{+0.23}$ & $-0.57_{-0.17}^{+0.16}$ & $-0.39_{-1.29}^{+0.62}$ & $-0.57_{-0.54}^{+0.48}$ \\
    $\alpha$ & 3.00 & $2.97_{-0.60}^{+0.86}$ & $3.15_{-0.73}^{+1.14}$ & $3.10_{-0.71}^{+1.20}$ & $4.07_{-1.41}^{+1.92}$ & $4.23_{-1.70}^{+2.94}$ & $5.11_{-2.23}^{+2.98}$ \\
    $\beta_{ \rm trans}$ & 1.00 & $1.47_{-0.51}^{+0.72}$ & $1.47_{-0.53}^{+0.87}$ & $1.43_{-0.5}^{+0.82}$ & $1.82_{-0.67}^{+0.75}$ & $1.51_{-0.81}^{+0.98}$ & $1.66_{-0.79}^{+0.87}$ \\
    $\gamma$ & 0.00 & $0.09_{-0.61}^{+0.48}$ & $0.11_{-0.46}^{+0.24}$ & $-0.01_{-0.81}^{+0.57}$ & $-0.61_{-0.91}^{+0.96}$ & $0.35_{-1.26}^{+0.69}$ & $-0.17_{-1.09}^{+0.76}$ \\
    $\log(\rho_0$/[M$_\odot$ pc$^{-3}$]) & $-0.22$ & $-0.44_{-0.47}^{+0.5}$ & $-0.64_{-0.28}^{+0.58}$ & $-0.43_{-0.57}^{+0.62}$ & $-0.51_{-0.59}^{+0.66}$ & $-0.69_{-1.24}^{+1.15}$ & $-0.61_{-0.90}^{+0.83}$ \\
    $\log(b)$ & 1.00 & $-0.63_{-0.25}^{+0.27}$ & $0.74_{-0.21}^{+0.22}$ & $0.65_{-0.30}^{+0.32}$ & $0.84_{-0.31}^{+0.22}$ & $0.96_{-0.51}^{+0.58}$ & $0.97_{-0.39}^{+0.53}$ \\
    
    \hline
  \end{tabular}
  \label{Tab:sphcoredresult}
\end{table}



\subsection{Effects of varying the number of samples stars and velocity errors}  \label{result:observables}
Effects of increasing the number of stars on the fitting with 4th-order moments can be observed by comparing the rightmost and middle columns in Figure \ref{fig:cornerplot_beta_gamma}. The posterior distributions demonstrate that increasing the sample size primarily reduces the uncertainty. This effect is particularly pronounced for $\beta$, where its uncertainty decreases by a factor of $\sim3$. Except for $\gamma$ in panel (e), the medians of other parameters remain largely unchanged when increasing the sample size from 500 to 5,000 stars. This suggests that, within this sample size range, the number of stars has a marginal influence on mitigating systematic biases, if there are any.




On the other hand, varying velocity error measurements show different behavior on the marginalized parameters. Notably, a dramatic shift emerges when we increase $\delta v_{\rm los}$ from 0.01 kms$^{-1}$ to  2.00 kms$^{-1}$ in the case of cored dSph mock galaxy, panels (d) and (e) in Figure \ref{fig:cornerplot_beta_gamma} and panels (a) and (b) in Figure \ref{fig:densityprofrecovsphcore}.
However, returning to the cuspy dSph case, the absence of such dramatic changes when increasing velocity errors suggests the existence of a threshold beyond which velocity errors no longer dominate the recovery of the density profile. Once this threshold is satisfied, increasing the number of stars becomes more critical than further reducing the velocity error. This finding has practical implications for observational surveys, as prioritizing larger sample sizes over extremely precise velocity measurements may yield more significant improvements in constraining the density profiles of dark matter halos, provided that the threshold is satisfied.

The differing responses of the two cases to the same velocity error are likely due to the velocity space being scaled to the total dispersion, $(\sigma_{\mathrm{los,}i}^2 + \delta v_{\mathrm{los,}i}^2)^{1/2}$, in the probability calculation. This implies that the smearing effects of observational errors on the LOSVD's shape depend not on the absolute value of the velocity error but rather on the ratio of $\sigma_{\textrm{los,}i}/\delta v_{\textrm{los,}i}$. Nonetheless, given the relatively small variations in $\sigma_{\textrm{los}}(R)$ in dSphs and the inaccessibility of the true $\sigma_{\textrm{los},i}$, the ratio $\sigma_{\textrm{los,global}}/\delta v_{\textrm{los,}i}$ can serve as a practical approximation for assessing whether the data is likely to produce biased parameter estimates. Here, $\sigma_{\textrm{los,global}}$ refers to the observed global velocity dispersion calculated from all sample stars, defined as $\sigma^2_{\textrm{los,global}} \equiv \sum_{i=1}^{N} (v_{\textrm{los,}i} - v_{\rm sys})^2/(N - 1)$.

To further investigate this velocity error dependency, we perform a similar fitting on the cored dSph mock galaxy data with a velocity error of 1.5 kms$^{-1}$. The results are $\gamma = 0.11^{+0.39}_{-1.83}$ and $\beta = -0.42^{+0.27}_{-0.27}$ at the 95\% confidence level, which closely resembles the results shown in panel (d) of Figure \ref{fig:cornerplot_beta_gamma} and is a significant improvement particularly for the median of $\gamma$ that is biased in panel (e). In this case, the ratio $\sigma_{\rm los,global} / \delta v_{\rm los}$ is 4.8. For comparison, this ratio is 3.6 in panel (e) of Figure \ref{fig:cornerplot_beta_gamma}, where the velocity error is set to 2.00 kms$^{-1}$, and 4.3 in panel (b). These results suggest that a ratio of $\sigma_{\rm los,global} / \delta v_{\rm los} \gtrsim 4$ is necessary to mitigate systematic biases introduced by velocity error measurements. This threshold aligns very closely with what is suggested by \citet{AmoriscoEvans2012}, who used a distribution function to trace symmetric deviations of LOSVDs from Gaussianity. This indicates that the threshold may be universal across different dynamical modeling approaches employing 4th-order moments rather than being specific to the Jeans or DF-based methods. Once the ratio reaches $\sigma_{\rm los,global} / \delta v_{\rm los} \approx 4$, however, there is little improvement left if we further increase the ratio. As expected, this requirement is significantly more restrictive than in the model relying only on the 2nd-order velocity moments as strong systematic bias is absent in panel (e) in Figure \ref{fig:densityprofrecovsphcore} and \ref{fig:cornerplot_beta_gamma} despite the relatively larger uncertainties. This resistance of the 2nd-order-only model to the velocity error measurements confirms a similar finding in \citet{Chang2021}.

\subsection{Results for the Ultra-faint mock galaxies}  \label{result:UFD}
\begin{figure}
  \centering
  \includegraphics[width=0.9\textwidth]{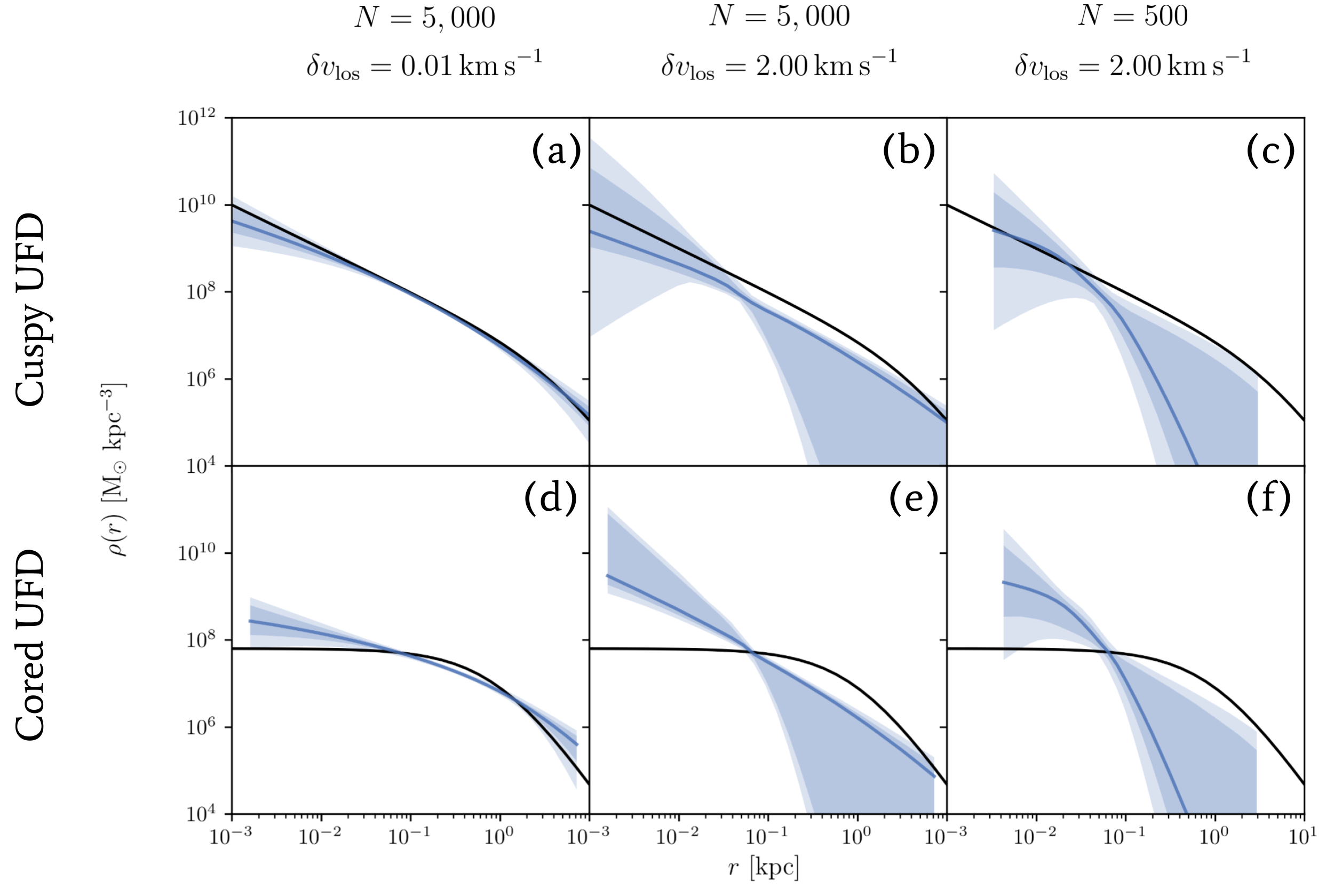}  
  \caption{The same as Figure \ref{fig:densityprofrecovsphcusp}, but for the UFD mock galaxies and only showing the results from fitting that incorporates 4th-order moments. Panels (a), (b), and (c) correspond to the cuspy UFD mock galaxy, while panels (e), (e), and (g) correspond to the cored one.}
  \label{fig:densityprofrecovufd}
\end{figure}

While the density profile recovery in dSph mock galaxies exhibits a relatively clear trend, the situation for UFD mock galaxies is far more challenging. Figure \ref{fig:densityprofrecovufd} illustrates the density profile recovery for UFD mock galaxies, following the same format as Figure \ref{fig:densityprofrecovsphcusp}, but exclusively presenting the model that incorporates the 4th-order moments.
\begin{table}[htbp]
  \centering
  \renewcommand{\arraystretch}{1.3} 
  \caption{Median and 68\% confidence level of the estimated parameters for the cuspy UFD mock galaxy}
  \begin{tabular}{cccccccc}
    \hline
    \multirow{3}{*}{Parameter} & \multirow{3}{*}{True value} & \multicolumn{2}{c}{$N = 5,000$} & \multicolumn{2}{c}{$N = 5,000$} & \multicolumn{2}{c}{$N = 500$} \\
                               &            & \multicolumn{2}{c}{$\delta v_{\rm los} = 0.01$ km s$^{-1}$} & \multicolumn{2}{c}{$\delta v_{\rm los} = 2.00$ km s$^{-1}$} & \multicolumn{2}{c}{$\delta v_{\rm los} = 2.00$ km s$^{-1}$} \\
                               &            & 2nd & 4th & 2nd & 4th & 2nd & 4th \\
    \hline  
    $\beta$ & 0.00 & $0.00_{-0.17}^{+0.14}$ & $-0.08_{-0.10}^{+0.08}$ & $0.07_{-0.23}^{+0.18}$ & $-0.50_{-3.01}^{+0.28}$ & $0.14_{-0.72}^{+0.41}$ & $-1.85_{-2.78}^{+1.81}$ \\
    $\alpha$ & 3.00 & $2.54_{-0.37}^{+0.88}$ & $2.35_{-0.25}^{+0.63}$ & $2.90_{-0.66}^{+1.22}$ & $2.27_{-0.21}^{+4.34}$ & $3.52_{-1.17}^{+3.06}$ & $5.16_{-2.10}^{+3.02}$ \\
    $\beta_{ \rm trans}$ & 1.00 & $0.50_{-0.19}^{+0.92}$ & $0.35_{-0.11}^{+0.49}$ & $0.83_{-0.44}^{+1.21}$ & $0.24_{-0.08}^{+1.91}$ & $1.17_{-0.74}^{+1.20}$ & $1.53_{-1.02}^{+0.97}$ \\
    $\gamma$ & 1.00 & $0.42_{-0.50}^{+0.55}$ & $0.47_{-0.21}^{+0.48}$ & $0.41_{-0.68}^{+0.44}$ & $0.42_{-0.28}^{+0.90}$ & $0.29_{-1.04}^{+0.66}$ & $0.67_{-1.24}^{+0.83}$ \\
    $\log(\rho_0$/[M$_\odot$ pc$^{-3}$]) & $-2.70$ & $-0.92_{-1.42}^{+1.40}$ & $-0.65_{-1.67}^{+0.93}$ & $-1.57_{-0.93}^{+1.68}$ & $-0.08_{-0.84}^{+0.70}$ & $-1.42_{-1.28}^{+1.82}$ & $-0.34_{-1.61}^{+1.27}$ \\
    $\log(b)$ & 1.00 & $1.17_{-0.57}^{+0.49}$ & $1.25_{-0.58}^{+0.48}$ & $1.30_{-0.55}^{+0.42}$ & $0.79_{-1.04}^{+0.79}$ & $1.31_{-0.60}^{+0.45}$ & $0.06_{-0.55}^{+1.28}$ \\
    
    \hline
  \end{tabular}
  \label{Tab:ufdcuspresult}
\end{table}

\begin{table}[htbp]
  \centering
  \renewcommand{\arraystretch}{1.3} 
  \caption{Median and 68\% confidence level of the estimated parameters for the cored UFD mock galaxy}
  \begin{tabular}{cccccccc}
    \hline
    \multirow{3}{*}{Parameter} & \multirow{3}{*}{True value} & \multicolumn{2}{c}{$N = 5,000$} & \multicolumn{2}{c}{$N = 5,000$} & \multicolumn{2}{c}{$N = 500$} \\
                               &            & \multicolumn{2}{c}{$\delta v_{\rm los} = 0.01$ km s$^{-1}$} & \multicolumn{2}{c}{$\delta v_{\rm los} = 2.00$ km s$^{-1}$} & \multicolumn{2}{c}{$\delta v_{\rm los} = 2.00$ km s$^{-1}$} \\
                               &            & 2nd & 4th & 2nd & 4th & 2nd & 4th \\
    \hline
    $\beta$ & 0.00 & $-0.16_{-0.21}^{+0.18}$ & $-0.11_{-0.12}^{+0.08}$ & $-1.82_{-3.50}^{+1.41}$ & $-0.42_{-4.43}^{+0.32}$ & $-1.46_{-3.68}^{+1.50}$ & $-3.11_{-2.18}^{+2.66}$ \\
    $\alpha$ & 3.00 & $2.85_{-0.59}^{+1.07}$ & $2.62_{-0.45}^{+0.85}$ & $3.29_{-0.90}^{+1.77}$ & $2.29_{-0.24}^{+5.20}$ & $5.11_{-2.22}^{+3.17}$ & $5.62_{-2.92}^{+2.76}$ \\
    $\beta_{ \rm trans}$ & 1.00 & $0.91_{-0.43}^{+1.12}$ & $0.45_{-0.16}^{+1.35}$ & $1.23_{-0.62}^{+1.05}$ & $0.28_{-0.11}^{+2.29}$ & $1.71_{-0.87}^{+0.86}$ & $1.54_{-1.13}^{+1.00}$ \\
    $\gamma$ & 0.00 & $0.26_{-0.65}^{+0.40}$ & $0.17_{-0.37}^{+0.48}$ & $-0.18_{-0.81}^{+0.58}$ & $0.61_{-0.19}^{+1.22}$ & $-0.50_{-0.95}^{+0.80}$ & $0.61_{-1.21}^{+0.97}$ \\
    $\log(\rho_0$/[M$_\odot$ pc$^{-3}$]) & $-1.19$ & $-1.39_{-0.65}^{+1.32}$ & $-1.07_{-1.22}^{+1.32}$ & $-1.27_{-0.66}^{+1.16}$ & $-0.52_{-0.94}^{+0.90}$ & $-1.13_{-0.78}^{+0.87}$ & $-0.31_{-1.10}^{+1.08}$ \\
    $\log(b)$ & 1.00 & $1.08_{-0.37}^{+0.45}$ & $1.39_{-0.33}^{+0.35}$ & $1.00_{-0.35}^{+0.41}$ & $0.36_{-0.44}^{+0.82}$ & $1.08_{-0.35}^{+0.48}$ & $-0.06_{-0.34}^{+0.90}$ \\
    
    \hline
  \end{tabular}
  \label{Tab:ufdcoredresult}
\end{table}
In panels (c) and (f) of Figure \ref{fig:densityprofrecovufd}, the model completely fails to recover the density profile. This stark failure highlights the difficulty in constraining the density properties of UFDs, even under the simplest assumptions of constant $\beta$ and spherical symmetry. In real observations, this task is further complicated by observational imperfections, such as binaries and interlopers, which are not accounted for in the model and might not be negligible in the small velocity dispersion of UFDs. This implies the significant challenges associated with placing meaningful constraints on UFD density profiles using currently available data. For UFDs, where the global l.o.s. velocity dispersion is of the order of $\sim4$ km s$^{-1}$ and the number of stars is typically $\sim100$, reliable inference of density profiles remains elusive.

Even when the number of stars is increased to 5,000 with velocity errors of 2 km s$^{-1}$, shown in panels (b) and (e), the model still largely struggles to trace the density profile accurately. The only partial success is observed in the inner regions of the cuspy case, where the model weakly detects an increasing density profile and recovers the density profile within a 68\% confidence level, albeit with significant uncertainties. This improvement is likely due to the extremely large $b$ parameter in the cuspy case, as discussed in Section \ref{subsec:dependencyonkappa} about the effects of $b$ on the density profile estimation. This generally poor performance is also closely tied to the ratio of velocity error to the global l.o.s. velocity dispersion. Specifically, the values of $\sigma_{\rm los, global}/\delta v_{\rm los}$ are 1.9 for the cuspy case and 1.7 for the cored case, both of which are well below the threshold of $\sigma_{\rm los, global}/\delta v_{\rm los} \approx 4$ identified for reliable recovery in dSph mock galaxies.

In panels (a) and (d), the density profiles are recovered within the 68\% confidence level for the cuspy case and the 95\% confidence level for the cored case. However, the median values are notably biased—leaning toward cored in the cuspy case and toward cuspy in the cored case. This apparent "midway" behavior suggests that the model struggles to make definitive distinctions between cores and cusps in UFDs, a challenge that is less pronounced in dSphs. Median values and 68\% confidence intervals from UFD mock galaxy fittings are summarized in Table \ref{Tab:ufdcuspresult} and \ref{Tab:ufdcoredresult}.

The primary cause of these difficulties in UFDs lies in their small l.o.s. velocity dispersion. To produce such small dispersions, the model requires putting low values for $\gamma$, $\rho_0$, and/or $b$, all of which are highly degenerate (as discussed in Section \ref{subsec:degeneracy3pars}). In the regime of small $\sigma_{\rm los}(R)$, variations in any parameters have small effects on the l.o.s. velocity dispersion profile. For instance, if we set $\rho_0 = 0.0125$ M$_\odot$ pc$^{-3}$, $b=10$, $\alpha = 3$, $\beta_{\rm trans}=1$, and $\beta = 0$, changing $\gamma$ from 0 to 1 only increases $\sigma_{\rm los}(R=a_*)$ from 1.2 km s$^{-1}$ to 3.6 km s$^{-1}$. If we change the scale density to $\rho_0 = 0.2$ M$_\odot$ pc$^{-3}$ while keeping the other parameters fixed, varying $\gamma$ from 0 to 1 increases $\sigma_{\rm los}(R=a_*)$ from 4.7 km s$^{-1}$ to 14.4 km s$^{-1}$. This characteristic occurs as well when $b$, rather than $\rho_0$, is the dominant factor driving the small $\sigma_{\rm los}(R)$. As a result, a wide range of parameter combinations can produce equally plausible solutions for the kinematic data, making it difficult to constrain the density profile of UFDs reliably.

Furthermore, obtaining a spectroscopic sample of 5,000 stars with precise l.o.s. velocity measurements in UFDs are unlikely to be feasible in the near future, even with the data released from the Subaru PFS survey. Obtaining proper motion data for stars in UFDs by the upcoming Nancy Grace Roman Space Telescope \citep{Sanderson2019} can be an option for mitigation to robustly constrain the dark matter distribution in these systems as demonstrated, for example, by \citet{Read2017} and \citet{Read2021}. By having 5D phase space, the velocity anisotropy parameter can be estimated directly, hence reducing much of the challenge currently lurking. Without such data, the limitations of the current models and data sets still continue to pose significant challenges to understanding dark matter distribution in UFDs.

\section{Discussion} \label{sec:discussions}
\subsection{The recovery of $\gamma$} \label{recoveryofgamma}
\subsubsection{Dependency on $\kappa_{\rm los}$ profile} \label{subsec:dependencyonkappa}
As one of the main objectives of this study, this subsection focuses on the recovery of $\gamma$. Notably, in the case of a cored halo density profile, the cuspy profile can be robustly excluded. For the cuspy case, however, significant probabilities remain within the cored parameter space. This difference can be attributed to the more pronounced non-Gaussian LOSVDs in the cored dSph mock galaxy particularly in the inner region, as indicated by $\kappa_{\rm los}(R)$ in the right panel in Figure \ref{fig:mockdatatrueprofiles}. These stronger deviations from Gaussian provide stronger signatures for the 4th-order moments to effectively rule out incorrect parameter regions.

\begin{figure}
  \centering
  \includegraphics[width=0.8\textwidth]{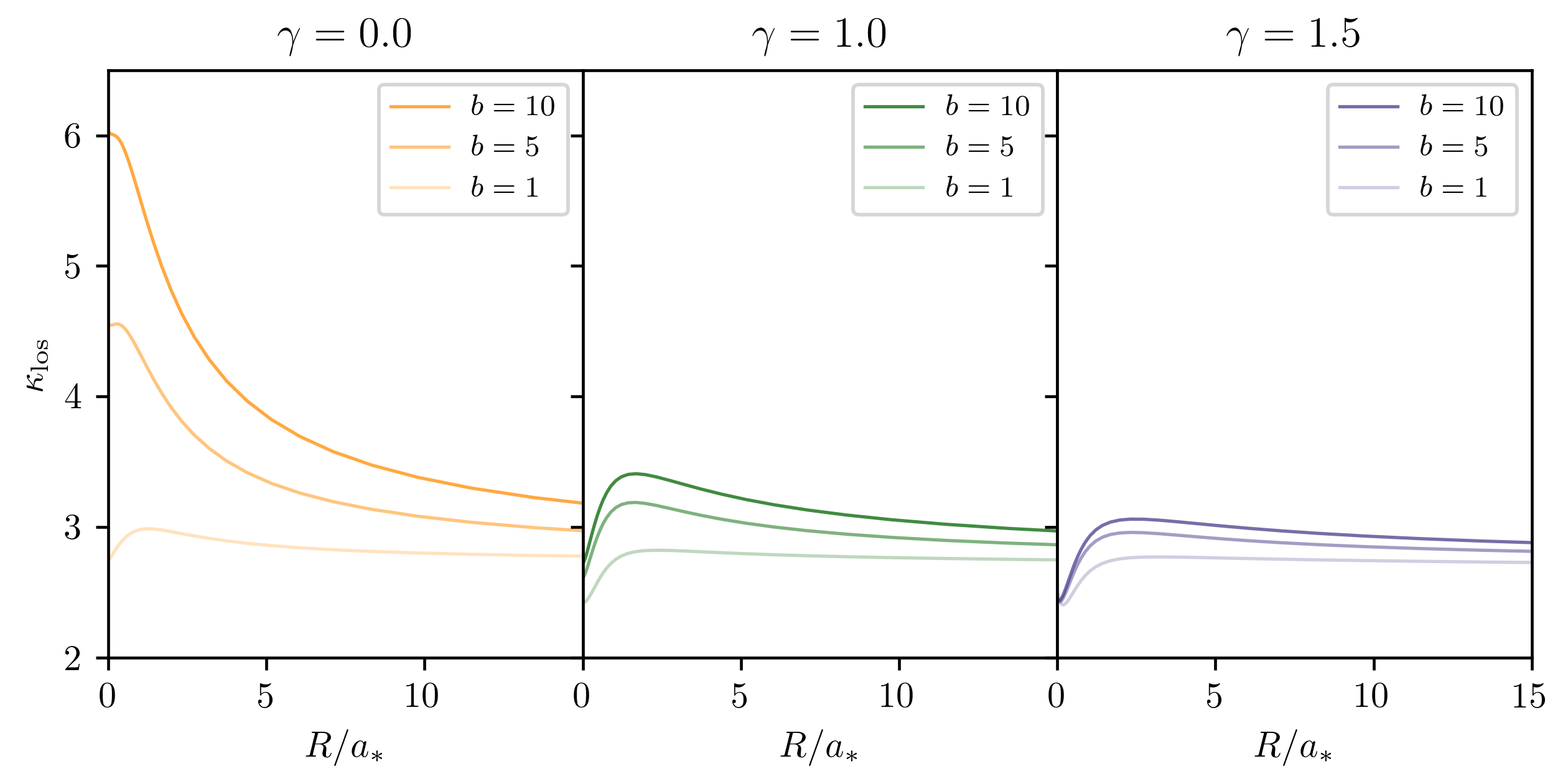}
  \caption{Different ranges of the l.o.s. kurtosis profile between DM halo with different value of $b$ (same $\gamma$ in one panel). The values are $\gamma=[0.0, 1.0, 2.0]$, respectively, from left to right. Meanwhile, the different line solidness in each panel denotes the change in $b$. The values are $b=[10, 5, 1]$, respectively, from the most solid line.}
  \label{fig:kappadiffgamma}
\end{figure}

To investigate deeper about the asymmetric posterior distribution of $\gamma$ in Figure \ref{fig:cornerplot_beta_gamma}, we plot l.o.s. kurtosis profiles by varying $\gamma$ and $b \equiv a_{\rm dm}/a_\ast$, i.e., the ratio between the scale lengths of dark and luminous matter. The results are shown in Figure \ref{fig:kappadiffgamma}. 
From left to right, the values of the dark matter inner density slope are $\gamma = [0.0, 1.0, 2.0]$, respectively.
In each panel, each line denotes $b = [1, 5, 10]$, respectively, from the lightest to the darkest, while the other parameters are fixed\footnote{$\beta=0$, $\alpha=3$, $\beta_{\rm trans}=1$, $\rho_0=0.064$ M$_\odot$ pc$^{-3}$}.
It is shown that in the case of the cuspier density profile, the range of $\kappa_{\rm los}(R)$ is narrower, being more concentrated around $\kappa_{\rm los} \approx 3$.
On the other hand, the range of $\kappa_{\rm los}(R)$ for a cored density profile is broader, reaching regions that are unlikely to be reached by a cuspy profile unless the other parameters have extreme values, a very large $b$ for example.
The narrow range of $\kappa_{\rm los}(R)$ in a cuspy halo explains the more rapid decrease in all posteriors of $\gamma$ for region $\gamma > 1$ compared to $\gamma<0$, despite the unphysical nature of the latter.
Another feature to note in Figure \ref{fig:kappadiffgamma} is the difference of $\kappa_{\rm los}(R)$ between a core and a cusp with the same value of $b$.
The gap in $\kappa_{\rm los}(R)$ between the cored and cuspy profile is more pronounced for a larger $b$. This allows the model to distinguish a core and a cusp more easily for a halo with a large $b$. The reason why a cored profile obtains more significant $\kappa_{\rm los}(R)$ boost when $b$ increases compared to a cuspy profile can be explained as follows.

The non-Gaussian LOSVD in a system with $\beta=0$ is a consequence of projection, i.e., integrating velocity distribution with different values of dispersion and star count. The smallest galactocentric distance along a certain $R$, hence $r_{\rm min}=R$, has the most star count in that line-of-sight. Therefore, the projected velocity dispersion along the line-of-sight at that point $\sigma_{\parallel}(r_{\rm min})$ has the most influence in determining $\sigma_{\rm los}(R)$.

Summing up velocity distributions along the l.o.s. ($r>r_{\rm min}$) with a lower star count ($\nu \propto r^{-5}$ outside the half-light radius) has different effects depending on the ratio $\sigma_{\parallel}(r > r_{\rm min})/\sigma_{\parallel}(r_{\rm min})$. If $\sigma_{\parallel}(r > r_{\rm min}) > \sigma_{\parallel}(r_{\rm min})$, the stars at $r>r_{\rm min}$ will populate the LOSVD's tail due to their broad distribution in the l.o.s. velocity space. In this case, the resulting LOSVD has more outliers than Gaussian, and consequently having $\kappa_{\rm los}>3$. Conversely, if $\sigma_{\parallel}(r > r_{\rm min}) < \sigma_{\parallel}(r_{\rm min})$, the majority of the stars at $r>r_{\rm min}$ will just be distributed within $\sigma_{\parallel}(r_{\rm min})$ and populating the "central" part of the LOSVD, resulting in a more flat-toped LOSVD with less outliers than Gaussian distribution.

\begin{figure}[h]
  \centering
  \includegraphics[width=0.8\textwidth]{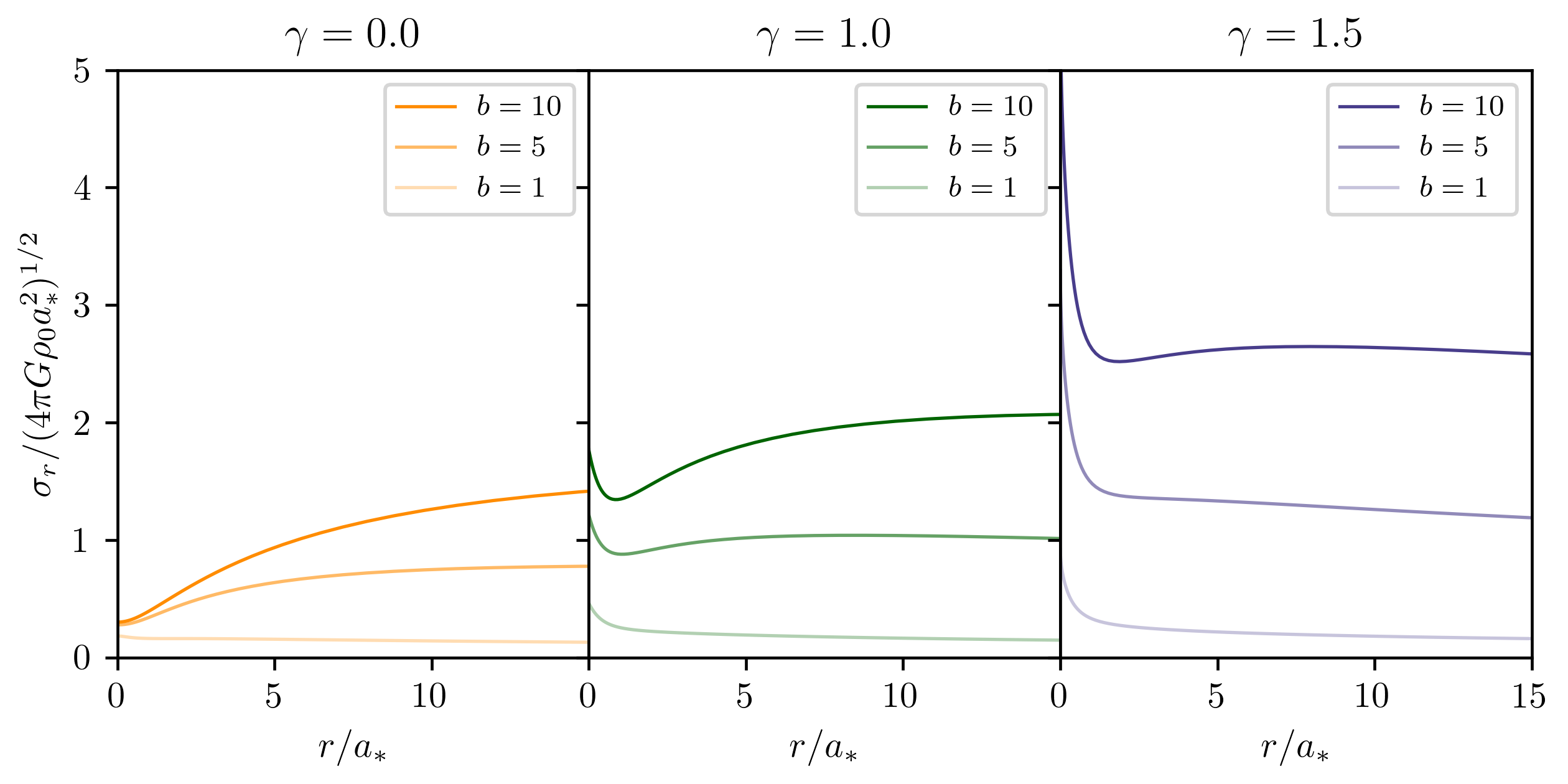}
  \caption{Normalized velocity dispersion profile in $r$ direction for dark halo with various $\gamma$ and $b$. The values are $\gamma=[0.0, 1.0, 1.5]$, respectively, from left to right as indicated above each panel. Different shades of line solidness in each panel denote the difference in $b$ where $b=[10,5,1]$, respectively, from the most solid line.}
  \label{fig:sigmar(r)diffgamma}
\end{figure}

Using $\sigma_{\parallel}(r>r_{\rm min})/\sigma_{\parallel}(r_{\rm min})$ to evaluate if the resulting LOSVD has fewer or more outliers than Gaussian is, of course, a simplification and aims only to provide a general sense because $\sigma_{\parallel}(r)$ along the line-of-sight might wind up and down. The LOSVD is a net product of all contributions from each point along the line-of-sight. To provide more illustration, let us consider galaxies having isotropic velocity anisotropy with, again, various $\gamma$ and $b$. Because of its isotropy, $\sigma_r(r)$ can represent any $\sigma_{\parallel}(r)$. Figure \ref{fig:sigmar(r)diffgamma} shows normalized $\sigma_r$ as a function of galactocentric distance $r$ scaled to the Plummer half-light radius $a_*$. The values of dark matter inner density slope are $\gamma = [0.0, 1.0, 1.5]$, respectively, from left to right and $b = [10, 5, 1]$, respectively, from the most solid one in each panel. The other parameters are $\beta_{\rm trans} = 1.0$ and $\alpha=3$.

The leftmost panel of Figure \ref{fig:sigmar(r)diffgamma} shows that because of its relatively low central density, a cored profile always exhibits a low velocity dispersion in the central part. This velocity dispersion generally increases with $r$, except in cases where $b \lesssim 1$. This characteristic, on the other hand, does not exist in a cuspy profile where the velocity dispersion in the central part is relatively larger than the core counterpart and generally decreases with $r$ before increasing again. The low central velocity dispersion with a large number of stars in a core halo provides a window for stars in the outer region with higher velocity dispersion to be outliers when they are integrated along the l.o.s. to obtain the LOSVD. Furthermore, because the enclosed mass at any $r$ is larger in a dark halo with a larger $b$, the velocity dispersion also increases more rapidly compared to dark halos with smaller $b$. As a result, outliers are more significant in a halo with a larger $b$, in agreement with the simple concept of $\sigma_{\parallel}(r>r_{\rm min})/\sigma_{\parallel}(r_{\rm min})$ ratio. This explains two characteristics; the first one is the reason for a considerable departure from $\kappa_{\rm los}(R) \approx 3$ for a cored dark halo with $b \gtrsim 5$ (the leftmost panel of Figure \ref{fig:kappadiffgamma}). The second is the reason for larger $\kappa_{\rm los}(R)$ \textit{everywhere} for a cored dark halo compared to the cuspy one with the same parameters (all panels in Figure \ref{fig:kappadiffgamma}).


\subsubsection{Degeneracy between $\gamma-\rho_0-b$} \label{subsec:degeneracy3pars}
The estimation of $\gamma$ is also challenged by a remaining degeneracy among the parameters of the dark matter distribution, namely $\gamma$, $b$, and $\rho_0$. It can be identified by the diagonally elongated contour in Figure \ref{fig:6pars_sph} and \ref{fig:6pars_sph2} in Appendix \ref{sec:6parsdistribution}. This issue may be an inherent limitation of parametric models unless one assumes a model-motivated specific relation between these parameters. The same degeneracy is also pointed out by \citet{Chang2021} even though the framework in their model reduces the degree of dark matter profile freedom by fixing the outer slope density profile $\alpha$ and the transition sharpness $\beta_{\rm trans}$ as well as does not set $\beta$ as a free parameter.

\begin{figure}[h]
  \centering
  \includegraphics[width=0.8\textwidth]{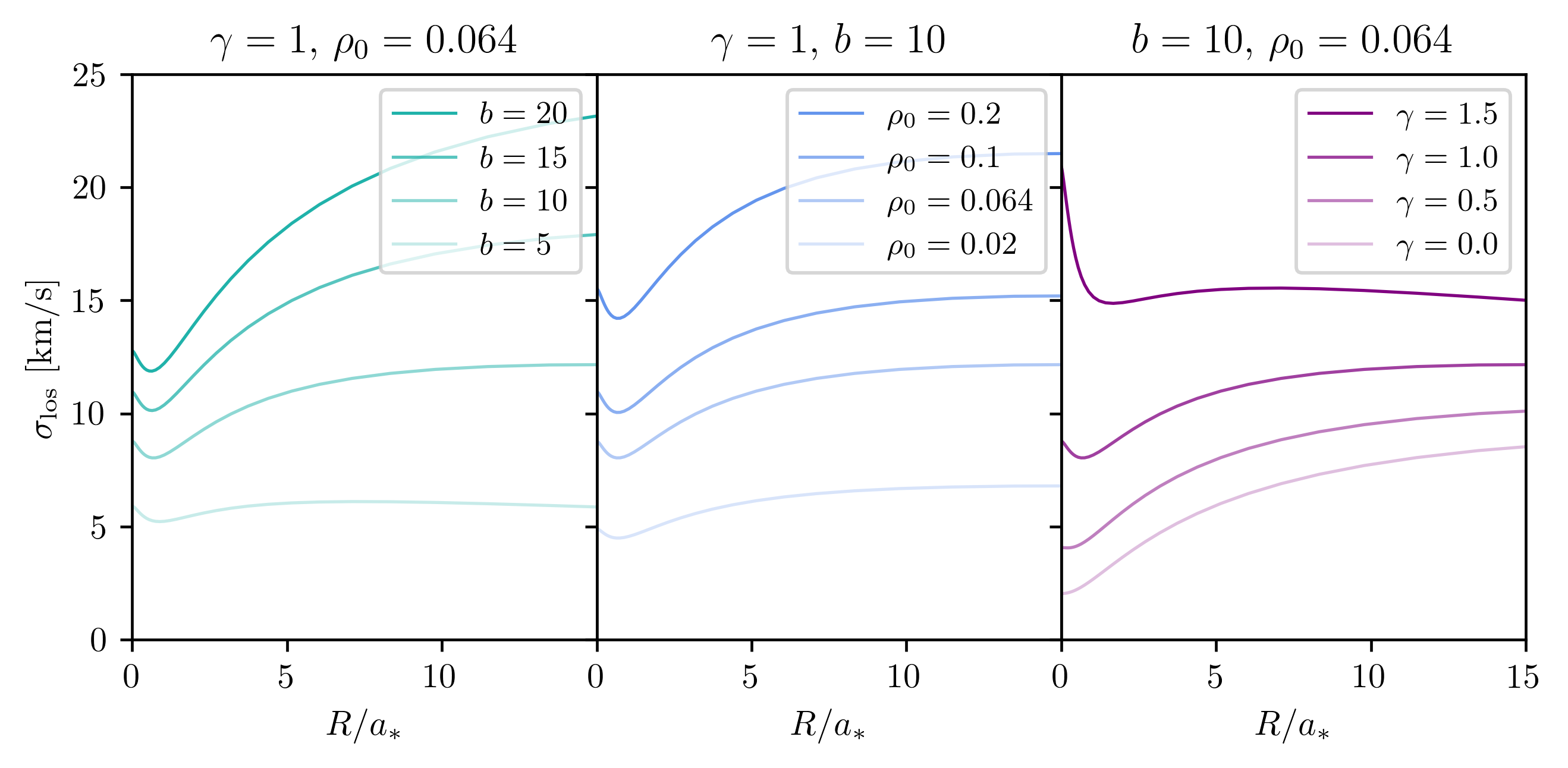}
  \caption{The l.o.s velocity dispersion profile of dark halos with a fixed $\gamma=0.0$ and $\rho_0 = 0.064$ M$_\odot~{\rm pc}^{-3}$ but varying $b=[20, 15, 10, 5]$ in the left panel, a fixed $b=10$ and $\rho_0 = 0.064$ M$_\odot~{\rm pc}^{-3}$ but varying $\gamma=[1.5, 1.0, 0.5, 0.0]$ in the middle panel, and a fixed $\gamma=1.0$ and $b=10$ while changing $\rho_0=[0.2, 0.1, 0.064, 0.02]$ M$_\odot~{\rm pc}^{-3}$ in the right panel, respectively, from the most solid line.}
  \label{fig:sigmalos(R)_gamma_b_degeneracy}
\end{figure}

To further explore the existence of the $\gamma-\rho_0-b$ degeneracy, we present l.o.s. velocity dispersion profiles in Figure \ref{fig:sigmalos(R)_gamma_b_degeneracy}. Each panel illustrates the effect of varying one parameter—$b$ (left panel), $\rho_0$ (middle panel), or $\gamma$ (right panel)—while keeping the other two fixed. Specifically, we use parameter values of $b=[20, 15, 10, 5]$, $\gamma=[1.5, 1.0, 0.5, 0.0]$, and $\rho_0=[0.2, 0.1, 0.064, 0.02]$ M$\odot$~pc$^{-3}$, starting with the most solid line. When fixed, we select $b=10$, $\gamma=1.0$, and $\rho_0=0.064$ M$_\odot$~pc$^{-3}$, as indicated in each panel. These panels illustrate that increasing $b$, $\gamma$, or $\rho_0$ produce similar behavior to the observed $\sigma_{\rm los}$ profile. It is clear that models utilizing only the 2nd-order moments struggle to distinguish this $\sigma_{\rm los}$ behavior from each other.

Additionally, Figure \ref{fig:kappadiffgamma} provides insight into why the 4th-order moments are unable to fully break this degeneracy. The figure shows that increasing $b$ (with $\gamma$ fixed) produces similar effects on the $\kappa_{\rm los}$ profile as decreasing $\gamma$ (with $b$ fixed). This similarity arises because both a larger $b$ and a smaller $\gamma$ correspond to a more gradual decline in the density profile outward. Furthermore, it is important to emphasize that the $\kappa_{\rm los}$ profile is independent of $\rho_0$, leaving the estimation of $\rho_0$ relies solely on the 2nd-order moment. Considering the remaining degeneracy even after incorporating the 4th-order moments, we suggest that a relation such as the mass-concentration relation in the CDM model \citep{Prada2012} or core mass-halo mass relation in the FDM model \citep{Schive2014b,Jowett2022} can help alleviate this challenge.

\subsubsection{Effects of varying the $R_{\rm max}$ of stars}
We examine the impact of the spatial distribution of stars by performing fittings with varying values of the maximum radial distance of the outermost member star, $R_{\rm max}$, while keeping the Plummer scale-length, $a_*$, and the dark matter scale-length, $a_{\rm dm}$, fixed. The ratio of these two scales is set to $b = a_{\rm dm}/a_* = 10$ for the cuspy dSph and $b = 4.5$ for the cored dSph mock galaxies. Two hard cuts are applied: $R_{\rm max} = 100 {\rm ~pc}=a_*$ and $R_{\rm max} = 500$ pc. The latter is selected to approximate the typical tidal radius of Milky Way dSph satellites, which is typically a few times the half-light radius. For each dSph mock galaxy, we generate 10 realizations by randomly selecting 5,000 member stars and assigning a negligible l.o.s. velocity error of $\delta v_{\rm los} = 0.01$ km~s$^{-1}$ to isolate the spatial limitation effects.

\begin{figure}
  \centering
  \includegraphics[width=0.8\textwidth]{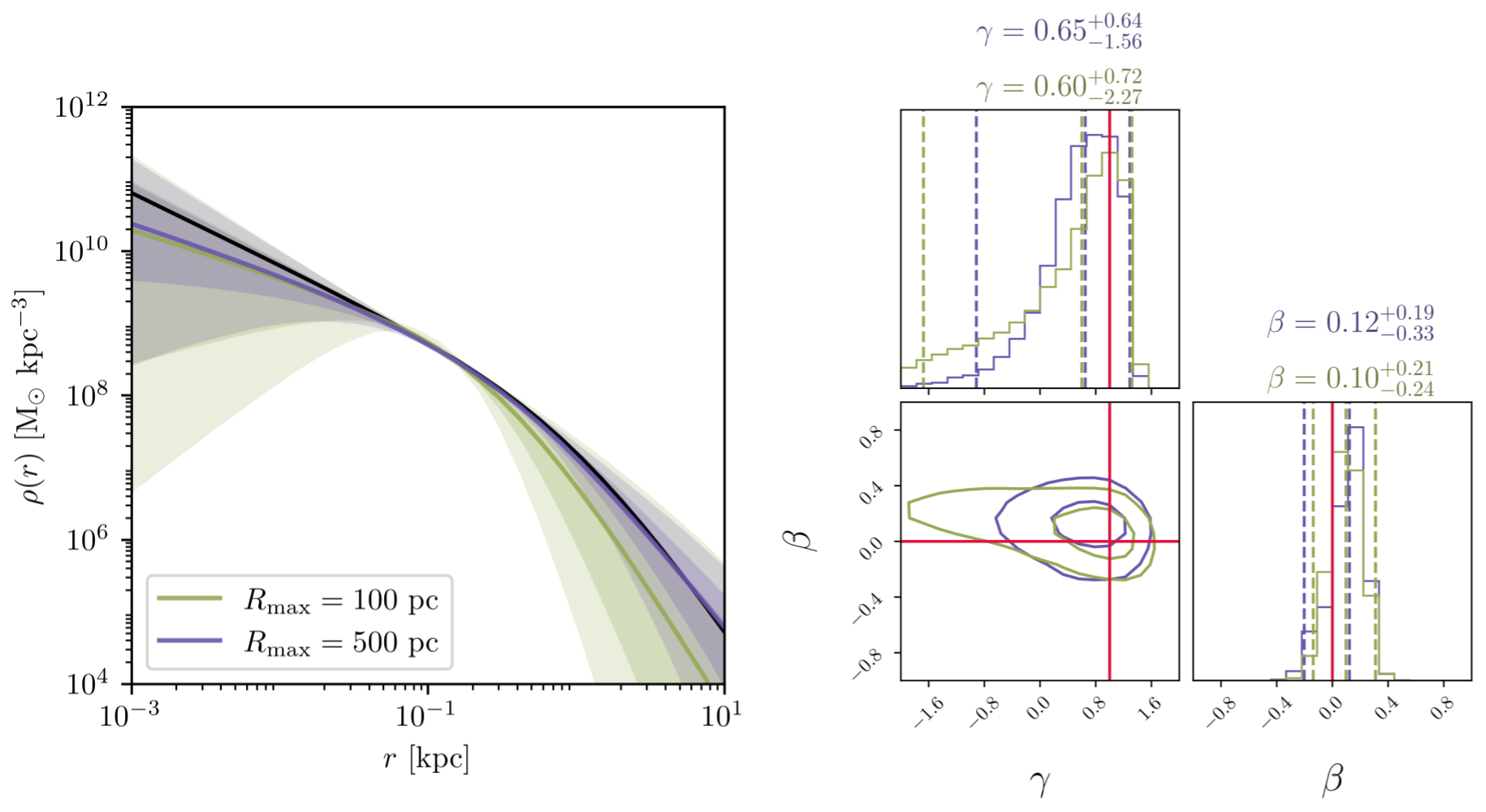}  
  \caption{Density profile recovery and posterior distributions for $\gamma$ and $\beta$ obtained from the fittings of 5,000 stars from the cuspy dSph mock galaxy and setting hard cuts on the tracer's spatial distribution at $R=5a_*=500$ pc (violet) and $R=a_*=100$ pc (green). All results are obtained by applying the model that incorporates 4th-order moments. In the left panel, the black solid line indicates the true profile and the darker (lighter) shade denotes the 68\% (95\%) confidence level. In the right panel, red solid lines denote the true values of each parameter, while the three vertical dashed lines indicate the median and 95\% confidence intervals. Contour lines in the 2-dimensional posterior distributions represent the 68\% and 95\% confidence levels.}
  \label{fig:varyingR}
\end{figure}

Results of the fitting are shown in Figure \ref{fig:varyingR} where green indicates $R_{\rm max}= 100$ pc and violet indicates $R_{\rm max}=500$ pc.
In the left panel, we show the density profile recoveries in the range of $1$ - $10^4$ pc. The range in this panel lies well beyond the availability of tracers in order to show the hard-cut effects of $R_{\rm max}$. The black solid line denotes the true profile, and the darker (lighter) shade denotes the 68\% (95\%) confidence level.
In the right panel, we show the posterior distributions for $\gamma$ and $\beta$ with the same color code. Red lines denote the true values and three vertical dashed lines denote the median and 95\% confidence level. Meanwhile, contour lines in the 2-dimensional posterior distributions represent the 68\% and 95\% confidence levels.

The results show that the true density profile is recovered within the 68\% confidence level in all scenarios.
However, there are increasing uncertainties and shifts in the median value as $R_{\rm max}$ is more restrictive. This occurs not only in the outer region, as expected, but also in the inner region.
Moreover, contours for $R_{\rm max} = 100$ pc reveal signs of a stronger degeneracy between $\gamma$ and $\beta$, suggesting that the spatial extent of tracers contributes to breaking the mass-anisotropy degeneracy.
This indicates that even though stars in the inner region play a dominant role in capturing the overall density profile trends, an extensive sample of up to a few $a_*$ is important to place useful limits on $\gamma$.

\subsection{The recovery of $\beta$}
\begin{figure}[h]
  \centering
  \includegraphics[width=0.8\textwidth]{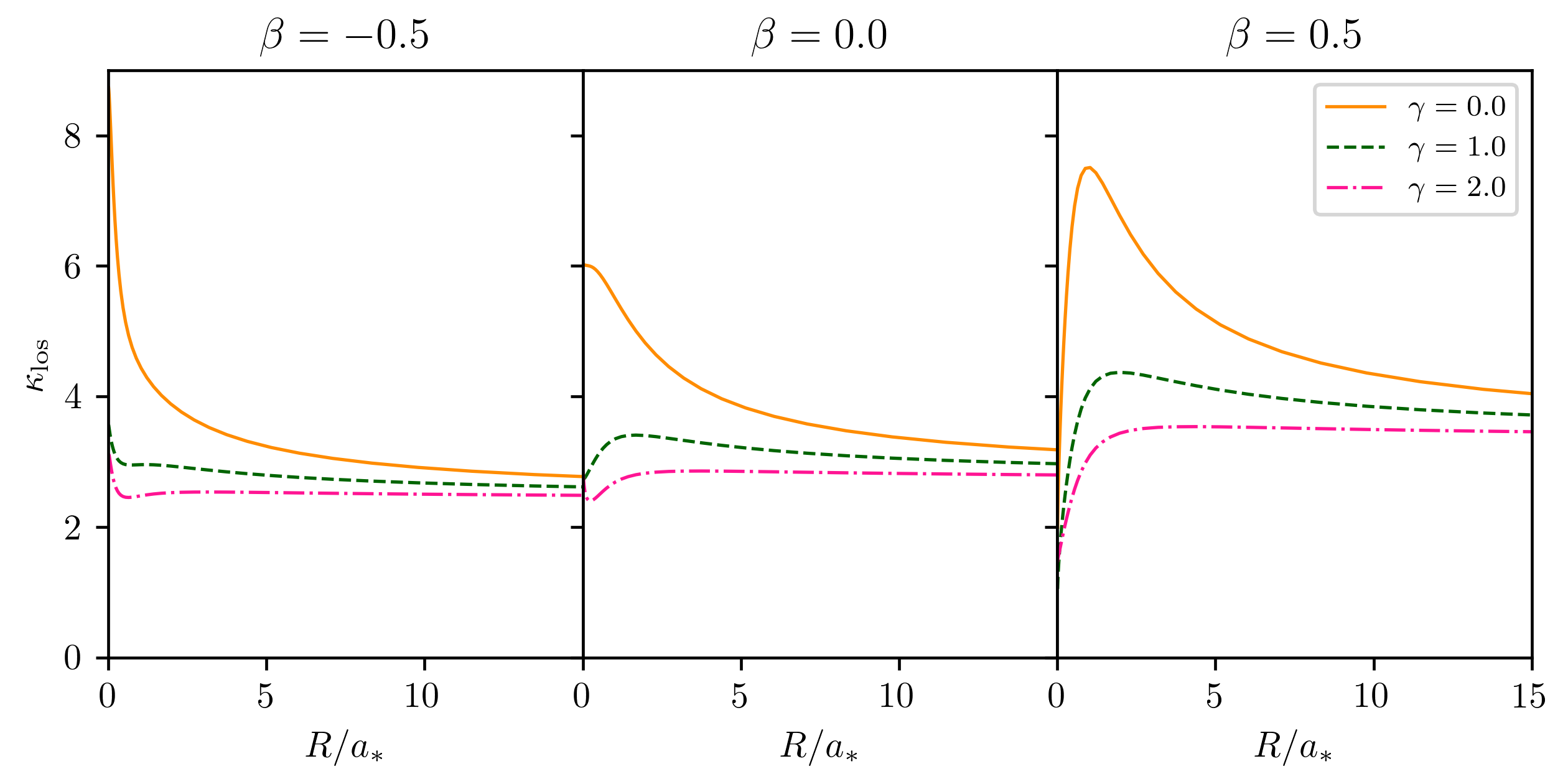}
  \caption{The line-of-sight kurtosis profile of dark halos with a fixed $b=10$ but different values of $\beta$ and $\gamma$. The values are $\beta=[-0.5, 0.0, 0.5]$, respectively, from left to right. Solid orange, dashed green, and dash-dotted pink lines denote $\gamma = [0.0, 1.0, 2.0]$, respectively.}
  \label{fig:kappalos(R)betavar_gvar}
\end{figure}

Apart from the estimation of $\gamma$, the estimation of $\beta$ is shown to be undisturbed by that situation of whether or not incorrect $\gamma$ can be ruled out. As long as $\sigma_{\rm los,global}/\delta v_{\rm los} \gtrsim 4$ is satisfied and the sample correctly traces the intrinsic kurtosis profile, $\beta$ is shown to be constrained within 95\% confidence level.
To investigate the main reason for the disentanglement between $\gamma$ and $\beta$, we provide Figure \ref{fig:kappalos(R)betavar_gvar} that shows the behavior of l.o.s. kurtosis profile when $\beta$ changes.
The leftmost panel corresponds to a system with tangentially biased orbits having $\beta= -0.5$, $\beta= 0.0$ for the middle panel, and $\beta= 0.5$ for the right panel.
From the leftmost panel, the values of $\beta$ are $-0.5$, 0.0, and 0.5, respectively.
Lines colored in orange, green, and pink denote different values of the dark matter inner density slope $\gamma = [0.0, 1.0, 2.0]$, respectively.
We fix the ratio of dark matter scale length to light scale length as $b=10$ in all panels.

From the previous section, we understand that increasing $\gamma$ results in smaller $\kappa_{\rm los}(R)$, which means more flat-topped LOSVDs with fewer outliers.
By contrast, as illustrated in Figure \ref{fig:kappalos(R)betavar_gvar}, an increase in $\beta$ leads to a decrease in $\kappa_{\rm los}(R)$ in the inner region while causing an increase in the outer region, according to the light and mass profiles adopted in our model.
This different response of $\kappa_{\rm los}(R)$ to changes in $\beta$ and $\gamma$ is the main reason for the disentanglement in the estimation of the two parameters.
Models relying solely on 2nd-order moments have no access to this distinction because increasing $\beta$ or $\gamma$ leads to a hardly distinguished increase in $\sigma_{\rm los}(R)$, particularly in $R < a_{\rm dm}$.
Moreover, unlike a large $\gamma$ that produces a narrow range of $\kappa_{\rm los}(R)$ as shown in Figure \ref{fig:kappadiffgamma}, $\beta$ has no regime that confines $\kappa_{\rm los}(R)$ in a similar restriction.
Even in the most subtle response of $\kappa_{\rm los}(R)$, that occurs when $\beta=0$, $\kappa_{\rm los}(R)$ is allowed to lie in a relatively wide range $2 \lesssim \kappa_{\rm los}(R) \lesssim 6$ as shown in the middle panel in Figure \ref{fig:kappalos(R)betavar_gvar}.
Thus, the model finds it less challenging to rule out incorrect values of $\beta$ than those of $\gamma$.
This sensitivity is the main reason why applying a higher-order moment analysis is able to place tight constraints on $\beta$.

In Figure \ref{fig:kappalos(R)betavar_gvar}, $\kappa_{\rm los}$ profiles in the outer region ($R \gg a_*$) monotonically increase as $\beta$ increases. This common trend can be explained by considering a line-of-sight in ($R \gg a_*$) region. The stellar density in this l.o.s. drops rapidly as $\nu(r) \propto r^{-5}$.
Therefore, the l.o.s. will be highly dominated by the stars located around the closest point of the l.o.s. to the galactic center $r_{\rm min} = R$.
In a system with $\beta > 0$, this point is filled by stars with $v_{\rm los} \approx v_{\rm sys}$ because the l.o.s. cuts stellar orbits almost perpendicularly, resulting in a strong-peaked LOSVD around $v_{\rm los} = v_{\rm sys}$ and, hence, a large $\kappa_{\rm los}$. Meanwhile, for a system with $\beta < 0$, this point is filled by a diverse value of $v_{\parallel}$ from stars of nearly circular orbits with different inclinations, resulting in a broad and flat-topped LOSVD.
A similar behavior is mentioned by \citet{BinneyTremaine2008} for a Hernquist density model for both light and mass profiles.
In addition, along any l.o.s., the velocity ellipsoid component parallel to the line of sight, $\sigma_{\parallel}(r)$, is always more aligned to $\sigma_r(r)$ with increasing $r$.
Because of this reason, high $v_{\rm los}$ stars in $r \gg r_{\rm min}$ populate the LOSVD tails in a system with radially biased orbits. However, it does not occur in a system with tangentially biased orbits because the l.o.s. cuts stars' orbits at approximately right angles.

The behavior of $\kappa_{\rm los}(R)$ in the inner region is more complex due to the more various interplay between the dark matter density profile and stellar density profile.
Instead of being heavily dominated by stars at $r_{\rm min}$, the integrated LOSVD is more dominated by stars lying in front of or behind $r_{\rm min}$ within the core Plummer profile due to the shallow stellar density gradient in that region.
Because of that, contribution to $v_{\rm los}$ in these regions is mainly obtained from the radial component of the star's orbital velocity.
If $\beta < 0$, most of the stars in those regions have $v_{\rm los} \approx v_{\rm sys}$, resulting in a strong spike around $v_{\rm los} = v_{\rm sys}$ in the LOSVD and a large $\kappa_{\rm los}$.
If $\beta > 0$, the LOSVD is filled by stars with various $v_{\rm los}$ from the radial velocity of stars with different galactocentric distances. This effect produces a broad, flat-topped LOSVD and a small $\kappa_{\rm los}$.

Even though the observed sensitivity of $\kappa_{\rm los}(R)$ in systems where $\beta \ne 0$ indicates a potential tendency for more robust parameter estimation compared to the $\beta = 0$ case, further investigation is still needed to confirm whether this sensitivity consistently leads to improved robustness in parameter estimation. \citet{Merrifield1990} suggests that the number of stars required to recover kurtosis in a heavy-tailed distribution is significantly larger than in a more thin-tailed distribution. This can intuitively be understood because to confirm that the distribution tails are strong, there must be a sufficient number of stars populating the relatively low probability regions. To conclude a thin-tailed distribution, by contrast, the stars just need to be absent. This is confirmed by \citet{Read2021} that apply four mass modeling methods to mock data and find that it is more challenging to recover the density profile and velocity anisotropy parameter in a system with $\beta(r)$ follows an Osipkov-Merrit profile \citep{Osipkov1979,Merritt1985} whose $\kappa_{\rm los}(R)>3$ across the entire range of radii.

It is also important to consider our assumption that $\beta(r)$ is constant. In reality, $\beta$ may vary with radius, as suggested by various studies. For example, \citet{loebman2018} simulate Milky Way-sized galaxies and observed a tendency that $\beta \sim 0$ in the innermost regions, transitioning to a radially biased $\beta \sim 0.7$ in the outer regions. Meanwhile, \citet{Klaudia2019} analyzes the Fornax dSph using a spherically symmetric Schwarzschild orbit superposition method, which does not require any assumption on the shape of $\beta(r)$. They find $\beta \approx 0$ near the center but gradually decrease with radius. These varied findings highlight the necessity of carefully studying systems with $\beta \ne 0$, as well as those with radially varying $\beta$ to obtain more accurate insight.

\section{Conclusions} \label{sec:conclusions}
We develop a dynamical model for dwarf spheroidal and ultrafaint dwarf galaxies using the 2nd-order and 4th-order spherical Jeans equations. In this model, we assume a constant velocity anisotropy parameter, $\beta$, with respect to galactocentric distance, and we consider the stellar mass to be negligible compared to the dark matter mass. The model is tested on four sets of mock galaxies to assess its performance in breaking the mass-anisotropy degeneracy and distinguishing between cored and cuspy density profiles. To examine the impact of observational limitations, we perform fits using different sample sizes, line-of-sight (l.o.s.) velocity measurement errors, and spatial distributions of stars. Additionally, we conduct fits that consider only the 2nd-order moments.

The results of our fitting demonstrate that incorporating 4th-order moments effectively disentangles $\beta$ from $\gamma$, partially breaking the degeneracy between these two parameters compared to the scenario where only 2nd-order moments are taken into account. We find that, in this model, a ratio of $\sigma_{\rm los,global}/\delta v_{\rm los} \gtrsim 4$ is necessary to mitigate most of the systematic biases associated with velocity errors. While this ratio plays the dominant role in controlling systematic bias, the number of stars has the greatest influence on the precision of the posterior distributions. Additionally, the results also suggest that the incorrect value of $\gamma$ can be ruled out more robustly as $\kappa_{\rm los}(R)$ deviates further from 3. Specifically, due to the nature of a cuspy density profile, which restricts $\kappa_{\rm los}(R)$ to a relatively narrow range compared to a cored profile, ruling out a cuspier profile than the true underlying one is less challenging than ruling out shallower density profiles.


While this model assumes spherical symmetry, real systems are often more complex and, thus, require a broader framework. Both observations and simulations indicate support for non-spherical models. Observationally, the light distribution in dSphs and UFDs is not spherically symmetric \citep{Munoz2018,Battaglia2022}. On the theoretical side, CDM simulations reveal that dark matter halos are generally non-spherical \citep{Allgood2006,Orkney2023}. In particular, studies of axisymmetric models have shown that distinct line-of-sight velocity dispersion profiles are projected along the major and minor axes, with notable differences between core and cusp profiles \citep{HayashiChiba2012}. Although the behavior of line-of-sight kurtosis in axisymmetric models remains unexplored, incorporating these features is expected to allow a more precise estimation of the true density profile in dwarf galaxies.

Despite these considerations regarding galaxy structure, our analysis in this study assumes idealized conditions, where the membership of each star can be determined with perfect accuracy. However, observational data often suffer from contamination by foreground or unbound stars \citep{BattagliaHelmiBreddels2013}. Although a detailed analysis of contamination effects is beyond the scope of this paper, it is worth noting that kurtosis-based models are generally more sensitive to such contamination than models using only 2nd-order moments, especially if the contaminations lie in the tails of the LOSVD \citep{Merrifield1990}. Additionally, the influence of binary stars needs to be carefully considered. While the contribution of binaries to inflating the velocity dispersion profile in dSphs is demonstrated to be minimal \citep{Minor2013,Spencer2017,Chema2023,Wang2023}, their effect could be more significant in UFDs and warrants careful treatment. A quantitative analysis of how unresolved binaries impact $\kappa_{\rm los}(R)$ will be an important focus of future work.

\begin{acknowledgments}
We thank the anonymous reviewer for constructive comments that helps revise the paper. We are also grateful to Laszlo Dobos, Carrie Filion, Evan Kirby, Alex Szalay, Rosie Wyse, and the Subaru/PFS GA working group for their useful discussion on the derivation and analysis of the LOSVD of dwarf satellites.
This work was supported in part by the MEXT Grant-in-Aid for Scientific Research  (No.~JP20H01895, JP21K13909, and JP23H04009 for KH, No.~JP18H05437, JP21H05448, and JP24K00669 for~MC.).
\end{acknowledgments}

%



\appendix \label{sec:Appendix}
\section{Kernels} \label{sec:kernels}

\subsection{Uniform Kernel}
To produce a distribution that is more boxy, having fewer outliers than a Gaussian distribution and a flat-topped peak, we can use the uniform kernel \citep{SandersEvans2020}
\begin{equation}
    K_+(y) = \frac{1}{2a}\textrm{, if $y<a$}.
\end{equation}
The error-convolved distribution is
\begin{equation}
    f_s(w) = \frac{b}{2a} \left[ \Phi \left( \frac{bw+a}{t} \right) - \Phi \left( \frac{bw-a}{t} \right) \right],
    \label{eq:fsuniform}
\end{equation}
with
\begin{equation}
    t^2 = 1 + b^2 \delta^2,
    \label{eqn:t}
\end{equation}
\begin{equation}
    b^2 = 1 + \frac{a^2}{k^2},
    \label{eqn:b}
\end{equation}
\begin{equation}
    k(a) = k_0 - (k_0 - k_\infty) \tanh{ \left( \frac{a}{a_0} \right) },
    \label{eqn:k}
\end{equation}
\begin{equation}
    \kappa = - \frac{2a^4}{15} \left( 1 + \frac{a^2}{3} \right)^{-2} + 3,
    \label{eq:kappalosuniform}
\end{equation}
where $\delta$ is the velocity error measurement and $\Phi(x)$ is the cumulative distribution function for the unit Gaussian PDF
\begin{equation}
    \Phi (x) = \frac{1}{\sqrt{2\pi}} \int_\infty^x \mathrm{d} t \exp \left( -\frac{t^2}{2} \right).
\end{equation}
The value for some appearing constants are $a_0=3.3, k_0=\sqrt{3}, k_{\infty} \approx 7/5$. This kernel can produce distributions with kurtosis lying in the range of $1.8 \leqslant \kappa < 3.0$.

\subsection{Laplacian Kernel}
To produce a distribution with more outliers than a Gaussian distribution, more heavy-tailed and a spikier peak, we can use the Laplacian kernel \citep{SandersEvans2020}
\begin{equation}
    K_+(y) = \frac{1}{2a} \exp{-\frac{y}{a}}.
\end{equation}
The error-convolved distribution is
\begin{equation}
    f_s(w) = \frac{b}{4a} \exp\left( \frac{t^2 - 2abw}{2a^2} \right) \erfc{\left( \frac{t^2-abw}{\sqrt{2}ta} \right)} + \\
    \frac{b}{4a} \exp \left( \frac{t^2 + 2abw}{2a^2} \right) \erfc{\left( \frac{t^2+abw}{\sqrt{2}ta} \right)},
    \label{eq:fslaplacian}
\end{equation}
\begin{equation}
    \kappa = \frac{12a^4}{(2a^2+1)^2}  + 3,
    \label{eq:kappaloslaplacian}
\end{equation}
with the same relation as Equations (\ref{eqn:t}), (\ref{eqn:b}), (\ref{eqn:k}) for $t$, $b$, $k$, respectively. The value for some appearing constants are $a_0=2.25, k_0=1/\sqrt{2}, k_{\infty,0}=1.08$. This kernel can produce distributions with kurtosis lying in the range of $3.0 < \kappa \leqslant 6.0$.

\section{Dependency on the prior of $\gamma$} \label{sec:priorgammadependence}

The solutions generated from Jeans modeling are not always guaranteed to be physically plausible. This limitation becomes particularly relevant when selecting the prior range for $\gamma$ because solutions with $\gamma < 0$ are unlikely to represent realistic physical conditions and can slow the convergence of MCMC chains. However, simply excluding $\gamma < 0$ from the prior range may unintentionally introduce boundary effects.

\begin{figure}
  \centering
  \includegraphics[width=0.65\textwidth]{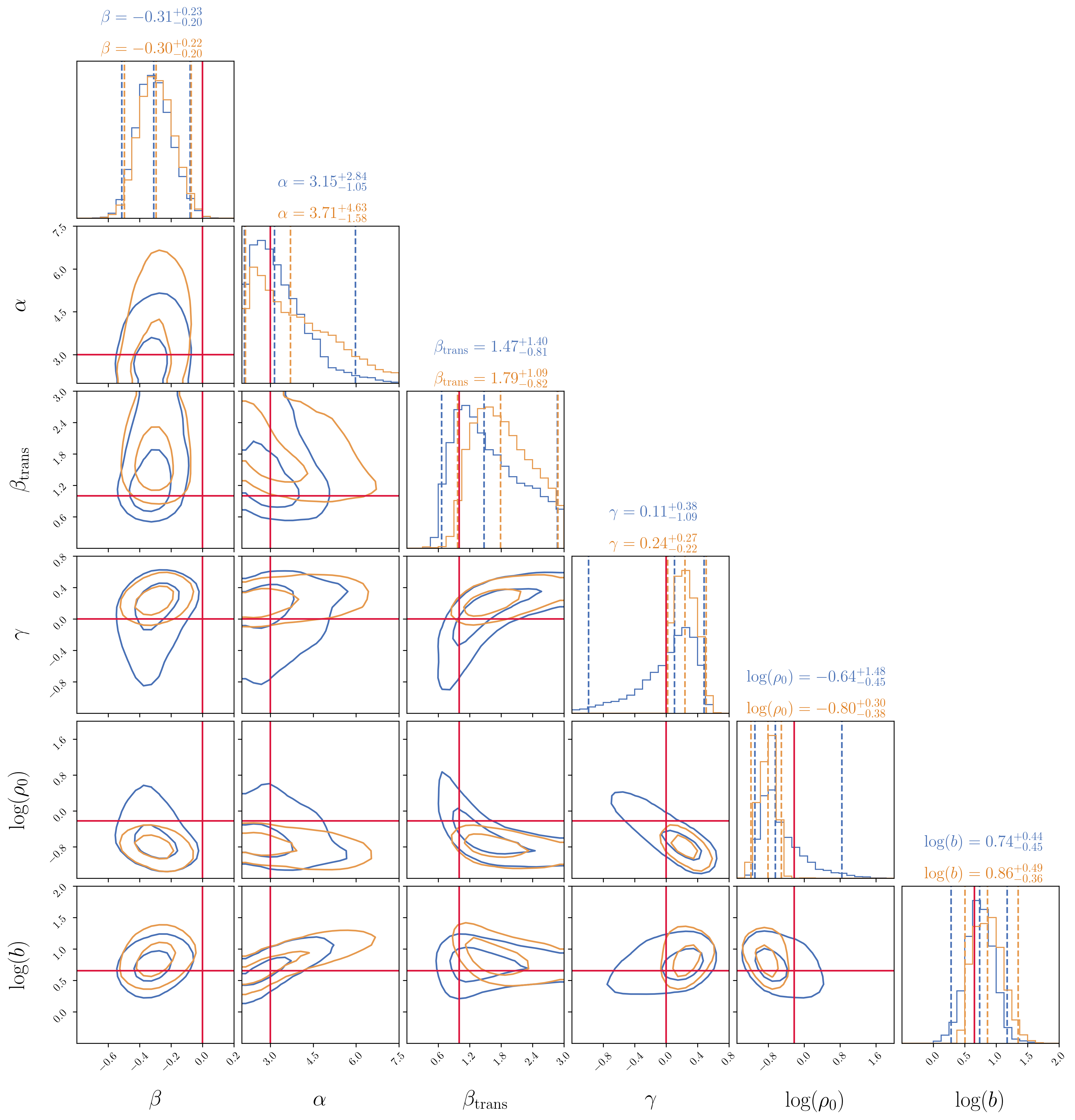}  
  \caption{Posterior distributions on $\gamma$ obtained from the fittings for the cored dSph mock galaxy by applying different ranges of prior. The blue posteriors correspond to a wide range of prior $-2.0 \leqslant \gamma \leqslant 2.0$, and the orange posteriors are obtained by assigning a narrower prior range of $0 \leqslant \gamma \leqslant 2.0$ that excludes unphysical values. The red solid lines denote the true value, while the three vertical dashed lines indicate the median and 95\% confidence intervals.}
  \label{fig:narrowgamma}
\end{figure}

To examine the impact of the prior choice on $\gamma$, we compare results using different prior ranges. Figure \ref{fig:narrowgamma} presents the posterior distributions of each parameter derived from fitting 5,000 randomly selected stars in the cored dSph mock galaxies. We apply two different ranges for the prior of $\gamma$: a broader range of $-2.0 \leqslant \gamma \leqslant 2.0$ (shown in blue) and a narrower range of $0 \leqslant \gamma \leqslant 2.0$ (shown in orange). The red lines indicate the true values. The estimated median values and 95\% confidence intervals for each case are also indicated above the corresponding column, as well as by three vertical dashed lines.

Except for the hard cut at $\gamma = 0$, Figure \ref{fig:narrowgamma} shows that the posterior distribution of $\gamma$ in the narrow-prior case closely resembles that of the broad-prior case. However, the other parameters must compensate for this restriction in $\gamma$ to maintain the fit to the data. These compensations propagate through the correlations between $\gamma$ and other parameters, leading to shifts, broadening, or artificial narrowing in their posterior distributions. This effect is, for example,  pronounced for $\rho_0$, where $\gamma - \rho_0$ correlation causes the exclusion of the true value. To minimize these prior-induced effects, we apply the broader prior $-2.0 \leqslant \gamma \leqslant 2.0$ to all our fittings.

\section{Posterior distribution for all parameters} \label{sec:6parsdistribution}
Figure \ref{fig:6pars_sph} and \ref{fig:6pars_sph2} presents the posterior distributions of all six free parameters in our analysis: $\beta$, $\alpha$, $\beta_{\rm trans}$, $\gamma$, $\rho_0$, and $b$, displayed in each panel from left to right. The left (right) column corresponds to the cuspy (cored) dSph mock galaxy. The number of stars and velocity errors are indicated on the left side for each row. All panels are obtained from a model that incorporates 4th-order moments. The red lines represent the true input values of each parameter, and the vertical dashed lines denote the median and 68\% confidence level. Correlations between $b$, $\gamma$, and $\rho_0$ appear in the two-dimensional distributions, as highlighted by the diagonally elongated contours.

\begin{figure}
  \centering
  \includegraphics[width=1.0\textwidth]{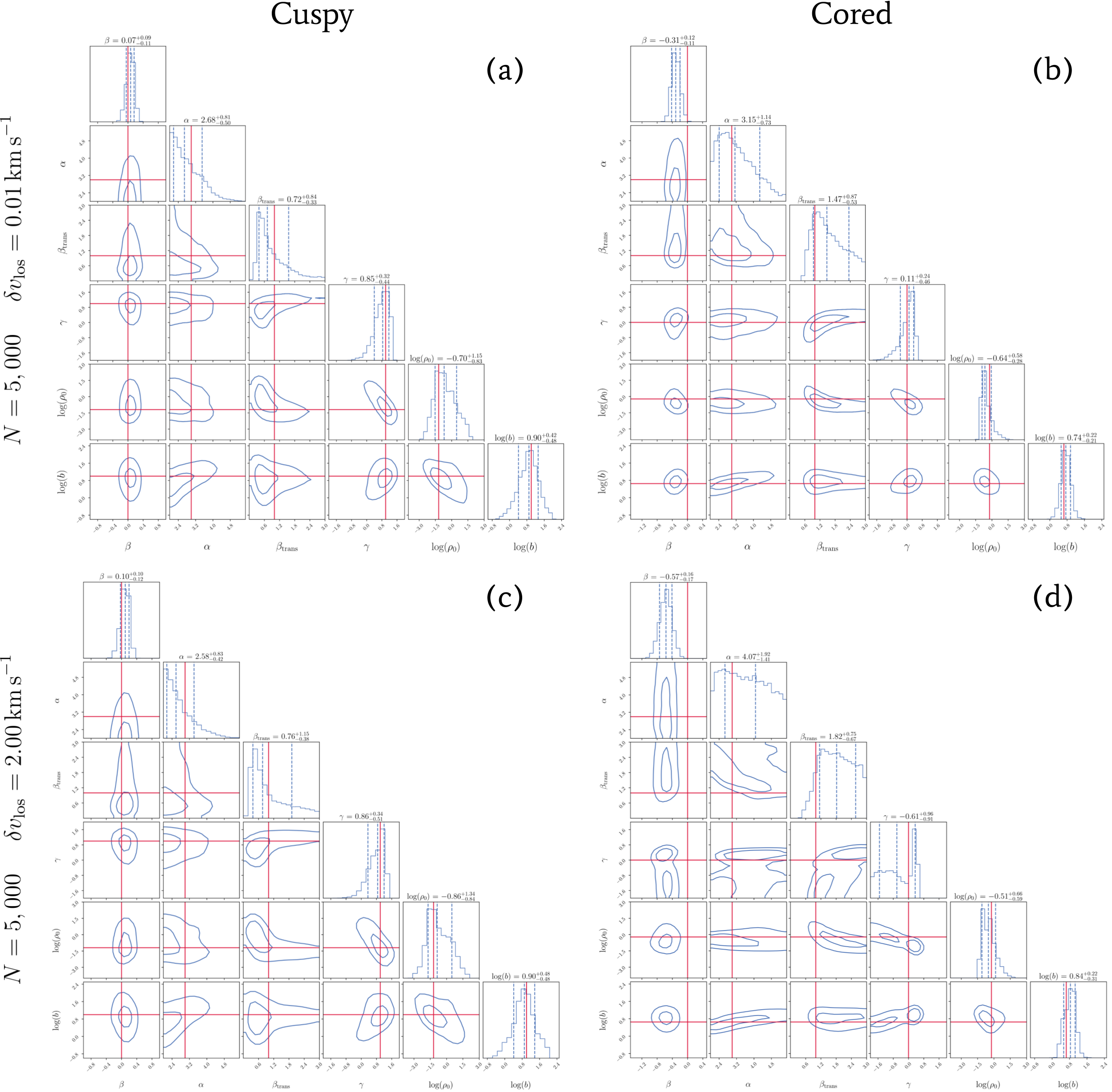}  
  \caption{Posterior distributions of all free parameters from the fittings of the cuspy dSph mock galaxy (left column) and the cored dSph mock galaxy (right column). The number of stars and velocity error used in each case are indicated on the leftmost side of each row. All results are obtained using the model incorporating 4th-order moments. Red solid lines indicate the true parameter values, while vertical dashed lines in the 1-dimensional posterior distributions represent the median and 68\% confidence intervals (note that these differ from the 95\% confidence intervals shown in Figure \ref{fig:cornerplot_beta_gamma}). Estimated median values and 68\% confidence intervals for each parameter are also indicated above the corresponding columns. Contour lines in the 2-dimensional distributions represent the 68\% and 95\% confidence levels.}
  \label{fig:6pars_sph}
\end{figure}

\begin{figure}
  \centering
  \includegraphics[width=1.0\textwidth]{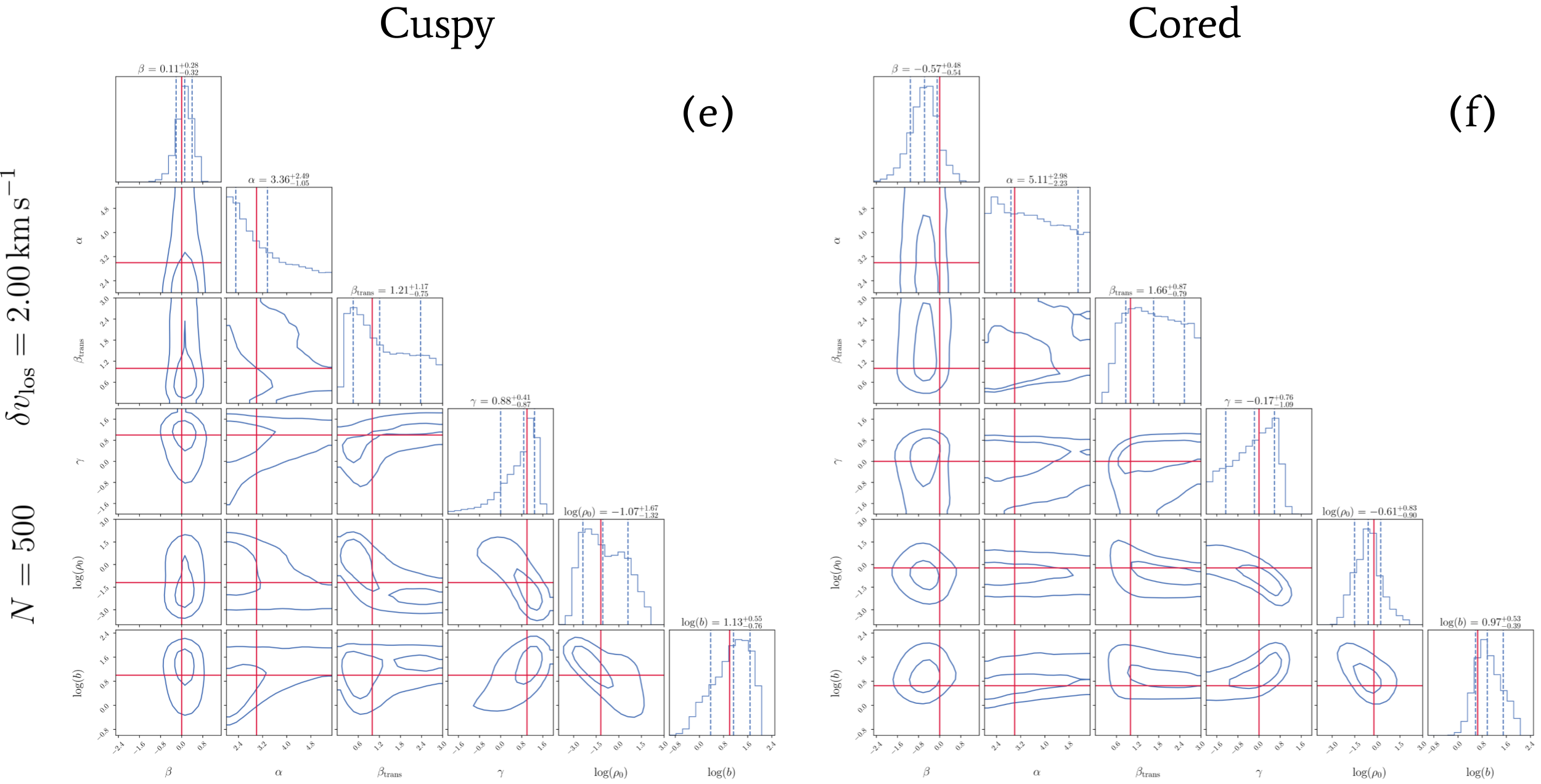}  
  \caption{Continuation of Figure \ref{fig:6pars_sph}.}
  \label{fig:6pars_sph2}
\end{figure}

\section{Standard deviation of the medians among realizations} \label{sec:realizationsdispersion}
Table \ref{Tab:mediandiffsph} summarizes the standard deviations of $\gamma$ and $\beta$ across multiple realizations for each fitting scenario applied to the dSph mock galaxies. These standard deviations provide a quantitative measure of the variability in the recovered parameters between different realizations.

\begin{table}[htbp]
  \centering
  \renewcommand{\arraystretch}{1.3} 
  \caption{Standard deviation of the medians across realizations from the results of dSph mock galaxies fitting.}
  \begin{tabular}{ccccccccc}
    \hline
    \multirow{3}{*}{Parameter} & \multirow{3}{*}{Density profile} & \multicolumn{2}{c}{$N = 5,000$} & \multicolumn{2}{c}{$N = 5,000$} & \multicolumn{2}{c}{$N = 500$} \\
                               &            & \multicolumn{2}{c}{$\delta v_{\rm los} = 0.01$ km s$^{-1}$} & \multicolumn{2}{c}{$\delta v_{\rm los} = 2.00$ km s$^{-1}$} & \multicolumn{2}{c}{$\delta v_{\rm los} = 2.00$ km s$^{-1}$} \\
                               &            & 2nd & 4th & 2nd & 4th & 2nd & 4th \\
    \hline
    \multirow{2}{*}{$\gamma$} & cuspy & $0.19$ & $0.09$ & $0.13$ & $0.24$ & $0.42$ & $0.31$ \\
                              & cored & $0.25$ & $0.18$ & $0.27$ & $0.71$ & $0.54$ & $0.50$ \\
    \hline
    \multirow{2}{*}{$\beta$} & cuspy & $0.08$ & $0.07$ & $0.09$ & $0.16$ & $0.30$ & $0.19$ \\
                              & cored & $0.14$ & $0.07$ & $0.16$ & $0.13$ & $0.63$ & $0.26$ \\
    
    \hline
  \end{tabular}
  \label{Tab:mediandiffsph}
\end{table}


\bibliography{sample631}{}
\bibliographystyle{aasjournal}



\end{document}